\documentclass[sigconf, nonacm]{acmart}

\usepackage{amsmath,amsfonts}

\usepackage{amssymb}
\usepackage{graphicx}
\usepackage{textcomp}
\usepackage{xcolor}
\usepackage{xspace}
\usepackage{graphics}
\usepackage{epsfig}
\usepackage{enumitem}
\newcommand{\Paragraph} [1] {\smallskip\noindent{\bf #1 }}
\usepackage{booktabs}
\usepackage{multirow}
\usepackage{bbding}
\usepackage{url}
\usepackage{tikz}
\usepackage{filecontents}

\usepackage{graphicx}
\usepackage{balance}  % for  \balance command ON LAST PAGE  (only there!)
\usepackage{subfig}
\usepackage{verbatim}
\usepackage{color,xcolor}
\usepackage{array}
\usepackage{bbding}
\usepackage{algorithm}
\usepackage[noend]{algpseudocode}
\usepackage{algpseudocode}
\newcommand\vldbdoi{XX.XX/XXX.XX}

\newcommand\vldbavailabilityurl{https://github.com/AiX-im/Sample-based-GNN}
\newcommand\vldbpagestyle{plain}

\newcommand{\system}{NeutronOrch\xspace}
\begin{document}
\title{NeutronOrch: Rethinking Sample-based GNN Training under CPU-GPU Heterogeneous Environments}

%\thanks{Qiange and Xin contributed equally. Yanfeng is the corresponding author.}
\vspace{-0.25in}
\author{Xin Ai, Qiange Wang, Chunyu Cao, Yanfeng Zhang, Chaoyi Chen, Hao Yuan, Yu Gu, Ge Yu}
\affiliation{%
  \institution{Northeastern University\country{China}}}
\email{{aixin0,wangqiange,chunyucao,chenchaoy,yuanhao}@stumail.neu.edu.cn, {zhangyf, guyu,yuge}@mail.neu.edu.cn}

\begin{abstract}

Graph Neural Networks (GNNs) have demonstrated outstanding performance in various applications. Existing frameworks utilize CPU-GPU heterogeneous environments to train GNN models and integrate mini-batch and sampling techniques to overcome the GPU memory limitation. 
 % This enables efficient model training on large graphs. 
In CPU-GPU heterogeneous environments, we can divide sample-based GNN training into three steps: sample, gather, and train. Existing GNN systems use different task orchestrating methods to employ each step on the CPU or GPU. After extensive experiments and analysis, we find that existing task orchestrating methods fail to fully utilize the heterogeneous resources, limited by inefficient CPU processing or GPU resource contention. 
%As a result, they limit the overall performance by either inefficient CPU processing or contention for GPU resources.

%In the early works, sampling and part of the data loading work was performed on the CPU, causing CPU contention. To solve the bottleneck caused by CPU contention, various GPU-led optimizations have been proposed one after another, bringing performance improvements while introducing GPU contention and CPU idle. 

%\zyf{The writing of the whole article needs significant improvement. too many small errors or typos would infuriate the reviewers. } 
% \zyf{is it too narrow if only push down the only one bottom layer? see Sec 4.1}
In this paper, we propose \system, a system for sample-based GNN training that incorporates a hotness-aware layer-based task orchestrating method and ensures balanced utilization of the CPU and GPU. \system decouples the training process by layer and pushes down the training task of the bottom layer to the CPU. This significantly reduces the computational load and memory footprint of GPU training. To avoid inefficient CPU processing, \system only offloads the training of frequently accessed vertices to the CPU and lets GPU reuse their embeddings with bounded staleness. Furthermore, \system provides a fine-grained pipeline design for the layer-based task orchestrating method, fully overlapping different tasks on heterogeneous resources while strictly guaranteeing bounded staleness. The experimental results show that compared with the state-of-the-art GNN systems, \system can achieve up to 11.51$\times$ performance speedup.

\end{abstract}

\maketitle
\pagestyle{\vldbpagestyle}

% \begingroup\small\noindent\raggedright\textbf{PVLDB Reference Format:}\\
% \vldbauthors. \vldbtitle. PVLDB, \vldbvolume(\vldbissue): \vldbpages, \vldbyear.\\
% \href{https://doi.org/\vldbdoi}{doi:\vldbdoi}
% \endgroup
% \begingroup
% \renewcommand\thefootnote{}\footnote{\noindent
% This work is licensed under the Creative Commons BY-NC-ND 4.0 International License. Visit \url{https://creativecommons.org/licenses/by-nc-nd/4.0/} to view a copy of this license. For any use beyond those covered by this license, obtain permission by emailing \href{mailto:info@vldb.org}{info@vldb.org}. Copyright is held by the owner/author(s). Publication rights licensed to the VLDB Endowment. \\
% \raggedright Proceedings of the VLDB Endowment, Vol. \vldbvolume, No. \vldbissue\ %
% ISSN 2150-8097. \\
% \href{https://doi.org/\vldbdoi}{doi:\vldbdoi} \\
% }\addtocounter{footnote}{-1}\endgroup
% VLDB block end %%%

%% do not modify the following VLDB block %%
%% VLDB block start %%%
\ifdefempty{\vldbavailabilityurl}{}{
\vspace{-0.0cm}
\begingroup\small\noindent\raggedright\textbf{PVLDB Artifact Availability:}\\
The source code, data, and/or other artifacts have been made available at \url{\vldbavailabilityurl}.
\endgroup
}
%%% VLDB block end %%%
\vspace{-0.2cm}
\section{Introduction}

%GNN背景
%GNN由传统DNN发展衍生而出，但传统DNN不能coverGNN任务，GNN有独特的图操作，要考虑都结构

Graph Neural Networks (GNNs) \cite{SURVEY_TNNLS_2021} are a novel class of Deep Neural Networks (DNNs) designed to process graph-structured data. They have demonstrated remarkable effectiveness across various graph applications\cite{GCN_iclr_2017,link_www_2021,link_neurips_2018,NMT_EMNLP_2017,Recom_kdd_22,REC_WWW_2019}.
GNNs aim to learn low-dimensional feature representations (i.e., embeddings) \cite{rep_ieee_2017} for each vertex in a graph and using the representations to graph-related tasks.
Recently, GPUs have been extensively used to accelerate GNN training owing to their high memory bandwidth and massive parallelism \cite{GNNGPU_PPOPP_2021,ROC_mlsys_2020}.
Graph learning systems like PyG \cite{PYG_corr_2019} and DGL \cite{DGL_CORR_2019} offer optimized GPU implementations and convenient programming interfaces to enhance both performance and usability of GNN training.

Considering the growing sizes of graphs in real-world applications, Full-graph GNN training that loads and trains the entire graph on the GPU is impractical due to the limited the GPU memory capacity \cite{hongtu_SIGMOD_2024,SAGA_SIGMOD_2022, MOSAIC_EUROSYS_2017, Marius++_arxiv_2022,SANCUS_VLDB_2022}. 
As a result, the sample-based approach for training GNNs has emerged as a promising solution. This approach performs computations on sampled subgraphs, allowing the training of large input graphs with limited GPU resources \cite{Graphsage_2017,GCNWS_SIGKDD_2018,DLG_TKDE_2022}.
%这里你说一下训练构成，简单但是准确，1）sampling-based training 把训练顶点划分batch。然后训练中，以每个batch的训练点为一个处理单元，首先通过采样操作，获取邻居子图（基于大概什么样的策略）。然后根据采样子图，收集涉及到的feature信息。 然后将采样子图和收集到的feature一起发送到GPU端执行训练。 并且将执行过程分为三个阶段
In this approach, the input data, including vertex features and graph structure, are stored in the host memory while the training process is offloaded to GPUs \cite{DGL_CORR_2019,distdgl_ieee_2020}. 
The sample-based GNN training divides the training vertices into multiple batches, treating each batch as the basic unit for training.
To train on a batch, the training process first samples the $K$-hop subgraphs for the training vertices, following a given neighbor sampling rule, then gathers the required vertex features based on the sampled subgraph and loads the subgraph and the involved vertex features onto the GPU. Finally, the GPU performs a forward and backward pass and updates the corresponding model parameters. 

This training mechanism, known as the \textbf{sample-gather-train} paradigm has gained widespread adoption in various GNN systems \cite{DGL_CORR_2019,PAGRAPH_SOCC_2020,GNNlab_EUROSYS_2022,DATATIRING_VLDB_2022,PYTDIRECT_VLDB_2021,Marius_OSDI_2021,Marius++_arxiv_2022,lazygcn_2020_neuips,TurboGNN_TOC_2023,wholegraph_SC_2022,quiver_2023_arxiv,FASTGcn_iclr_2018,DSP_PPOPP_2023,ducati_sigmod_2023,GNNautoscale_icml_2021,refresh_arxiv_2023,legion_atc_2023}. These systems decouple the three steps and deploy them on different computation devices to achieve high performance. 
%Depending on the resources they mainly utilized, each step can work in a CPU-centric or GPU-centric mode.  
Early systems \cite{PAGRAPH_SOCC_2020,DGL_CORR_2019,lazygcn_2020_neuips,FASTGcn_iclr_2018,Marius++_arxiv_2022} employ graph sampling and feature gathering on CPUs. These systems store the graph data in the CPU memory, perform CPU-based graph sampling, and collect the required features on the CPU side before transferring them to GPUs. This approach allows the GPUs to be dedicated solely to training tasks. On the other hand, some other systems \cite{Nextdoor_eurosys_2021,GNNlab_EUROSYS_2022,DATATIRING_VLDB_2022,DGL_CORR_2019,TurboGNN_TOC_2023,wholegraph_SC_2022,ducati_sigmod_2023,DSP_PPOPP_2023,legion_atc_2023} leverage GPU-accelerated graph sampling. They store the graph topology in either GPU or pinned CPU memory and perform the sampling on GPUs, relieving CPUs from the heavy random memory access. Moreover, recent systems \cite{PAGRAPH_SOCC_2020,GNNlab_EUROSYS_2022,DATATIRING_VLDB_2022,TurboGNN_TOC_2023,wholegraph_SC_2022,legion_atc_2023,DSP_PPOPP_2023,ducati_sigmod_2023} also employ GPU-based feature gathering to optimize overall performance. They store the vertex feature completely or partially (with a caching mechanism) on GPUs, converting the heavy host-GPU data communication and CPU memory access to efficient GPU global memory access. The GPU-based graph sampling and GPU-based feature caching can be used jointly to optimize performance.

%The training step is usually accelerated using GPUs. 
%Therefore, various task orchestrating strategies exist in sample-based GNN systems for the sampling-gathering-training process. 

\begin{figure}[!t]
  \centering
  \includegraphics[width=\linewidth,page={1}]{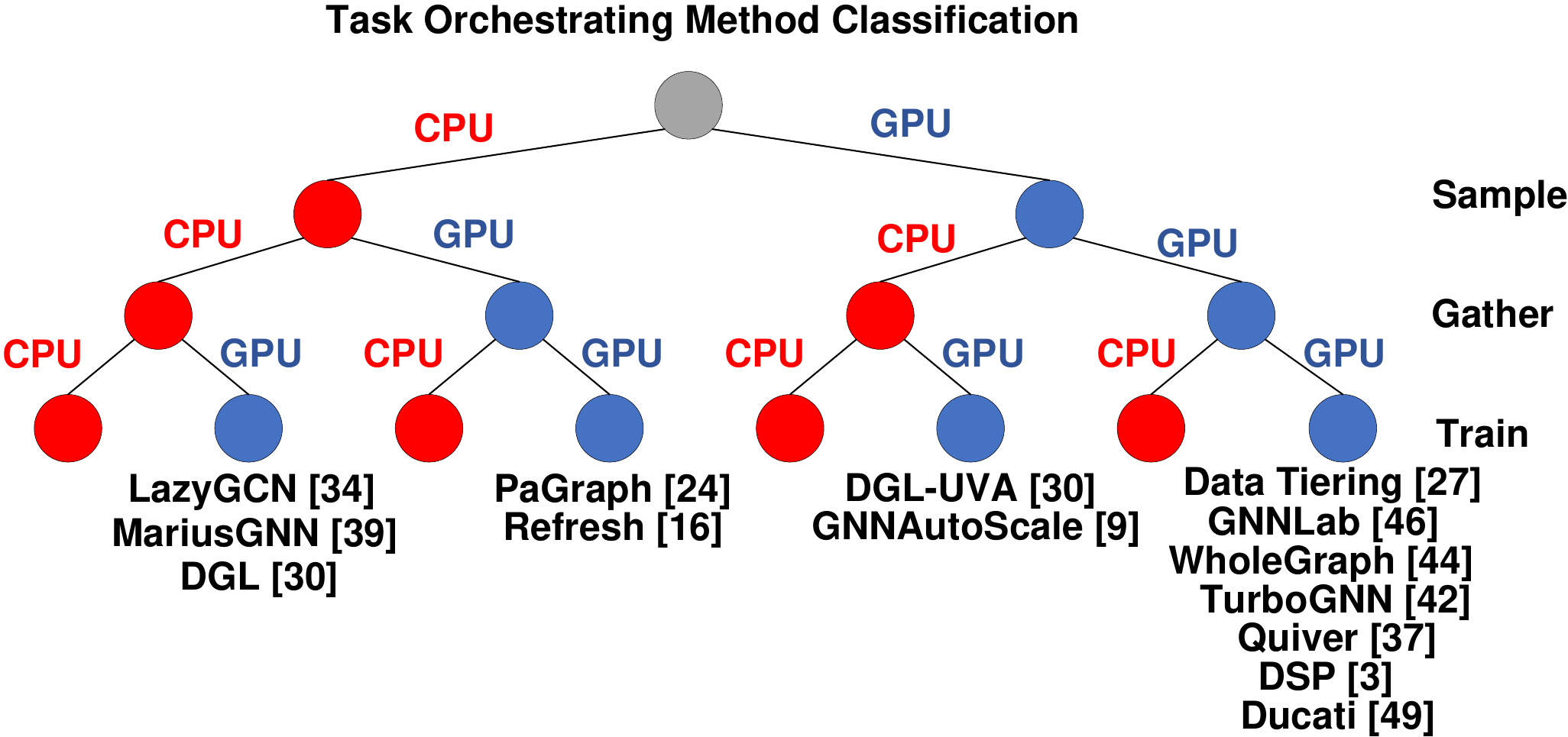}
  \vspace{-0.8cm} %调整图片与上文的垂直距离
  \caption{Classification tree of existing GNN systems and their task orchestrating methods. }
     \vspace{-0.4cm}
  \label{fig:system_classification}
\end{figure}

Figure \ref{fig:system_classification} provides an overview of existing GNN systems and their task orchestrating methods using a 3-layer binary tree. Each path from the root to the leaf represents a specific task orchestrating method, where the CPU is selected when traversing the left child, and the GPU is selected when traversing the right child. 
%Among the possible eight methods, since performing training on the CPU is not a wise option under CPU-GPU heterogeneous environments, the existing task orchestrating methods can be summarized into four categories.
Since performing training on the CPU is not a efficient option when GPU is equipped, the existing task orchestrating methods on CPU-GPU heterogeneous framework contains only four categories.

Having conducted intensive experiments and analysis, we find that existing task orchestrating methods that decouple the computation based on the step does not fully utilize CPU-GPU heterogeneous resources due to inefficient CPU processing or GPU resource contention. Assigning one or two
steps to the CPU may cause inefficient CPU processing to become
a bottleneck and leading to long waiting time for the GPU training process to receive the input training data.
In contrast, assigning two or more steps to the GPU may result in GPU resource contention although GPU parallel processing can significantly enhance the performance of each single step. This is because the training data, cached features, and graph topology need to be simultaneously stored within the constrained GPU memory while the computation kernels of sampling and training are completed for GPU cores. 
% The step-based task orchestrating methods fail to achieve adequate and balanced CPU and GPU utilization. 

\begin{figure}[!t]
  \centering
  \includegraphics[width=\linewidth,page={1}]{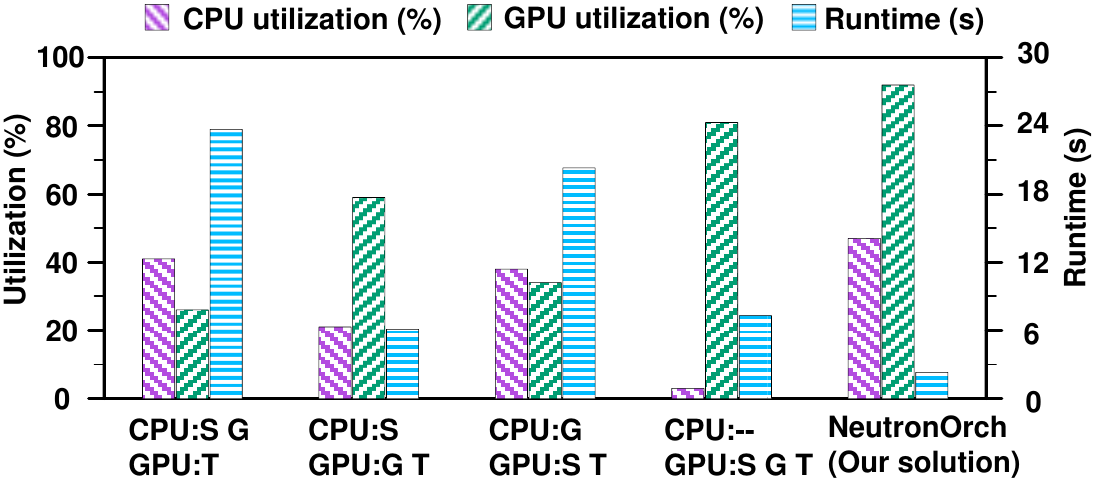}
  \vspace{-0.6cm} %调整图片与上文的垂直距离
  \caption{Comparison of resource utilization and per-epoch execution time for different task orchestrating methods. The S, G, and T represent the sample, gather, and train steps of sample-based GNN training.}
     \vspace{-0.5cm}
  %\zyf{font style is times new roman, but the font style in Fig. 1 is Arial or . You should make the fond style consistent in all figures.}
  %\zyf{the font style is not consistent with others}} %\zyf{S,D,T should be explained in fig or captain}}
  \label{fig:utilization}
\end{figure}

% In contrast, Figure \ref{fig:system_classification} illustrates that many systems \cite{DATATIRING_VLDB_2022,GNNlab_EUROSYS_2022,wholegraph_SC_2022,TurboGNN_TOC_2023,DGL_CORR_2019,PAGRAPH_SOCC_2020} rely on GPUs to handle two or three steps. Although GPU-based optimizations can enhance the performance of each step, there is a risk of GPU resource contention if many steps are deployed on the GPU. 
%which also prevents us from overlapping the computation of different batches. 
%These two cases, \textbf{(1) CPU contention with GPU idle}, and \textbf{(2) CPU idle with CPU contention}. 
%GNN training usually overlaps different batches for concurrent execution, and the sampling kernel and training kernel may compete for GPU cores. Therefore, 
%We summarize the major problems of existing computation orchestrating strategies into two types: \textbf{(1) GPU contention but CPU idle}, \textbf{(2) GPU wait caused by CPU contention}. 
%Most applications developed for heterogeneous systems are more suitable for the GPU than for the CPU. Only considering the device suitability when scheduling the applications overloads the GPU and leaves the CPU idle. On the other hand, overusing the CPU and making the GPU wait only makes the situation worse. 
%Figure \ref{fig:utilization} displays four task orchestrating methods that we configure in DGL \cite{DGL_CORR_2019}. 
To illustrate this, we implement the four task orchestrating methods in DGL \cite{DGL_CORR_2019} and compare their performance on a 3-layer GCN model with the Reddit dataset \cite{GCN_iclr_2017}. We evaluate the GPU utilization, CPU utilization, and per-epoch runtime as shown in Figure \ref{fig:utilization}. 
We can observe that the existing task orchestrating methods do not effectively utilize CPU-GPU heterogeneous resources due to either unbalanced resource utilization or low overall resource utilization, resulting in sub-optimal performance.

% \ccy{(Modified: 1. GNNLab 2. Alignment: Sample Gather Train 3. Last row horizontally aligned)}

%\wqg{To balance the resource utilization and improve the overall performance, we reposition the roles of CPU and GPUs in the pipeline of sampling-based GNN training, and propose a new computation orchestrating method. In traditional GNN systems, the CPU is often excluded from computation tasks due to its low parallelism. However, we observe that using the CPU can help alleviate the contention of GPU, especially, reduce the transmission overhead between CPU and GPUs. Different from existing GNN training methods that place each step into a single device, we decouple the computation by layers and distribute the computation to both CPU and GPU.}
%In traditional GNN systems, the CPU is often excluded from computation tasks due to its low parallelism. However, we observe that using the CPU can help alleviate the resource contention of the GPU and decrease the total transfer amount.

In this work, we rethink the roles of CPU and GPU in sample-based GNN training and propose a hotness-aware layer-based task orchestrating method to optimize performance. Unlike existing methods that allocate each step to a single device, we decouple the training task by layers and employ the computation of each sub-task (sample-gather-train) to a single device. This is based on our observation that the multiple hops of neighbor sampling cause most of the training resources (especially the memory) consumption to be in computations from the bottom. Offloading computation of the bottom layer to the CPU can significantly reduce the computation and memory requirement of the GPU. Furthermore, the volume of CPU GPU communication can be significantly reduced by transferring computed feature embeddings instead of raw features. Considering offloading the computation of a complete GNN layer to the CPU may make the CPU processing a new bottleneck if not carefully optimized, we propose a hotness-aware embedding reusing method that reduces CPU computation by periodically computing the embeddings of frequently accessed vertices (i.e., hot vertices) and reusing them across batches.
To sustain the correctness of training with embedding reuse, we extend bounded staleness \cite{ssp_2013_nips,SANCUS_VLDB_2022,ST_ICML_2018,Marius_OSDI_2021} processing to guarantee that the reused embedding is within a given stale constraint. Moreover, to further improve performance, we propose a super-batch pipelined training that combines $k$ adjacent batches into a super-batch and controls the staleness of hot embeddings among super-batches, overlapping GPU and CPU computation tasks while strictly and efficiently implementing bounded staleness.

By integrating the above techniques, we propose \system, a sample-based GNN system that can efficiently utilize the resources of heterogeneous CPU-GPU environments. 
% We evaluate the performance of \system using two popular GNN models, namely GCN \cite{GCN_iclr_2017} and GraphSAGE \cite{Graphsage_2017}, across five real-world graphs. Experimental results demonstrate that, compared with three state-of-the-art GNN systems, DGL \cite{DGL_CORR_2019}, Pagraph \cite{PAGRAPH_SOCC_2020}, and GNNlab \cite{GNNlab_EUROSYS_2022}, \system achieves up to 8.64X, 7.11X, and 2.02X speedups, respectively. 
% 
We make the following contributions.

\begin{itemize}[leftmargin=*]
    \item We provide a comprehensive analysis of resource utilization issues associated with the task orchestrating methods for sample-based GNN systems on GPU-CPU heterogeneous platforms.
    
    \item We propose a hotness-aware layer-based task orchestrating method that effectively leverages the computation and memory resources of the GPU-CPU heterogeneous system.
    
    \item We propose a super-batch pipelined task scheduling method that seamlessly overlaps different tasks on heterogeneous resources and efficiently achieves strict bounded staleness.
    %simultaneously reduce the amount of computation and transfer and increase the value of CPU computing power.
    %\item Based on the GPU-centric, CPU-assisted computing framework, we design and implement \system, a CPU-GPU balanced sample-based GNN system with a new caching strategy and well-designed pipeline \wqg{I think we need to "rethink" this, isn't new caching our main method ?}
\end{itemize}

We evaluate the performance of \system using three popular GNN models, GCN \cite{GCN_iclr_2017}, GraphSAGE \cite{Graphsage_2017}, and GAT \cite{GAT_ICLR_2018}. 
Experimental results demonstrate that, compared with five state-of-the-art sample-based GNN systems, DGL \cite{DGL_CORR_2019}, Pagraph \cite{PAGRAPH_SOCC_2020}, GNNlab \cite{GNNlab_EUROSYS_2022}, DSP \cite{DSP_PPOPP_2023}, and GNNAutoScale \cite{GNNautoscale_icml_2021}, \system achieves speedups of up to 11.51$\times$, 9.72$\times$, 2.43$\times$, 1.63$\times$, and 9.18$\times$, respectively.

\section{Background}
%GNN; mini-batch training; existing framework and their challenges; summary;
\label{sec2}

\subsection{Graph Neural Networks}

% \zyf{lowercase $k$ is a specific layer, uppercase $K$ is the total number of layers}
%首先是传统GNN的训练过程，按层计算，聚合邻居。其次，通过图的规模不断增大的特点以及图的幂律分布特性，引出Mini-batched gnn training
Graph-structured data serves as the input for Graph Neural Networks (GNNs), where each vertex or edge is associated with a high-dimensional feature vector.
A typical GNN model comprises multiple layers that compute a low-dimensional embedding for each vertex. 
Each layer contains an aggregation phase and an update phase \cite{GNNlayer-2020}. 
For example, in a GNN with $L$ layers, during the aggregation phase of layer $l$, each vertex $v$ combines its neighbors' embedding vectors at the $l - 1$ layer with its own embedding vector to generate the aggregation result $a_{v}^l$ using an aggregation function:
\begin{align}
    a_{v}^{l} &=AGGREGATE(h_u^{l-1} | \forall {u} \in {N_{in}(v)} \cup  \{v\})
 \label{form:agg_forward}
\end{align} 
\noindent where ${N_{in}(v)}$ represents the incoming neighbors of vertex $v$, $h_v^{l}$ represents the node embedding vector of vertex $v$ at l-th layer, and $h_v^0$ is the input feature of vertex $v$. The aggregate functions can be \texttt{sum}, \texttt{average}, \texttt{max/min}, etc. Next, during the update phase, each vertex computes its output embedding vector $h_{v}^l$ by applying an update function to the aggregation result $a_{v}^l$:
\begin{align}
    h_v^{l}&=UPDATE(a_{v}^{l}, W^{l})
 \label{form:update_forward}
\end{align}
After $L$ layers, each vertex’s feature vector becomes a low-dimensional embedding of its neighbors up to $L$ hops away. These embeddings can be utilized for performing a wide range of tasks, such as node classification and link prediction.

\begin{table}[tbp]
	\centering 
	\caption{\label{tab:test}Notations}
 \vspace{-0.1in}
  \scalebox{0.75}{
	\begin{tabular}{lcr}
		\toprule
		Notation & Description  \\
		\midrule
        $\mathbf{a}_v^{l}$ & aggregation results vector of vertex $v$ at the $l^{th}$ layer \\
        $\mathbf{h}_v^{l}$ & embedding vector of vertex $v$ at the $l^{th}$ layer \\
		$\mathbf{A_i}$ & Adjacency matrix of the $i^{th}$ iteration \\
		$\mathbf{\hat{A}_i}$ & Adjacency matrix after symmetric normalization of the $i^{th}$ iteration \\
		$\mathbf{W}_i^{l}$ &   Weight matrix at the $l^{th}$  layer of the $i^{th}$ iteration\\
		$\mathbf{H}_i^{l}$ & Embedding matrix at the $l^{th}$ layer of the $i^{th}$ iteration  \\
		$\mathbf{Z}_i^{l}$ & Input matrix to activation function at the $l^{th}$ layer of the $i^{th}$ iteration \\
        $\nabla \mathcal{L}$ & Gradient of the loss function \\
		$\nabla_{\mathbf{Z}^{l}} \mathcal{L}=\delta^{l}$ &  Gradient matrix of the loss $\mathcal{L}$ with respect to $\mathbf{Z}^{l}$ at the $l^{th}$ layer \\
		$\nabla_{\mathbf{W}^{l}} \mathcal{L}$ & Gradient matrix of the loss $\mathcal{L}$ with respect to $\mathbf{W}^{l}$ at the $l^{th}$ layer  \\
        $g(W_i)$ & Gradient of the loss function with respect to $W$ at the $i^{th}$ iteration \\
		$\mathcal{L}$& Loss of the GNN \\
        $I$ & Number of iterations in an epoch or number of batches \\
        $L$ & Number of layers of GNN model \\
		$\eta$& Learning rate \\
		$\epsilon$& Staleness bound \\
		\bottomrule
	\end{tabular}
 }
 \vspace{-0.2cm}
\end{table}

\begin{figure}[t]
  \centering
  \includegraphics[width=\linewidth,page={1}]{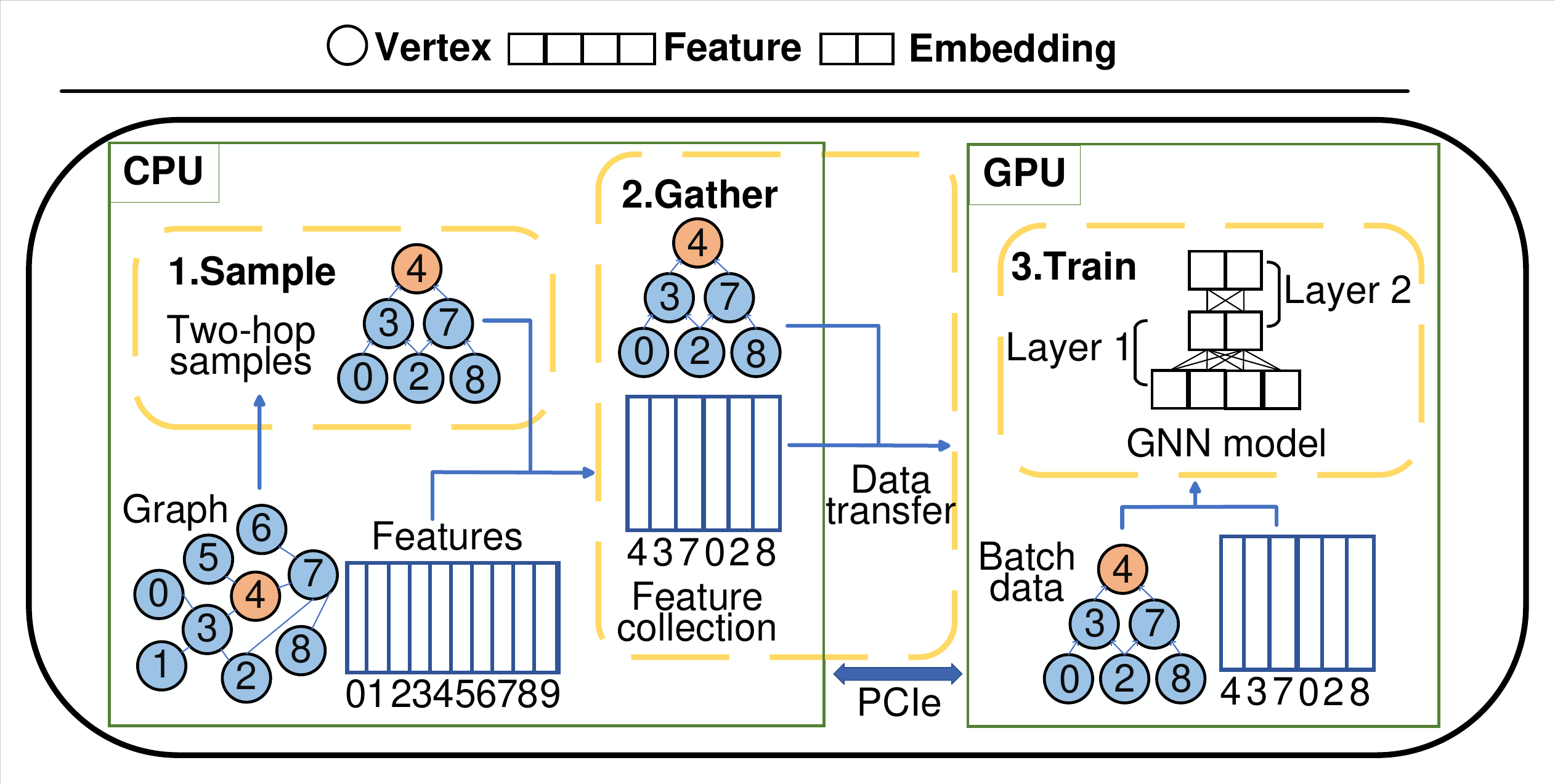}
   \vspace{-0.4cm}
  \caption{An example of sample-based GNN training for a two-layer GNN, where vertex 4 is the vertex with a ground-truth label for training.}
   \vspace{-0.4cm}
  \label{fig:Sample}
\end{figure}

\begin{algorithm}[t]
\caption{Sample-based mini-batch GNN training (per epoch)} \footnotesize 
\label{alg:sampleflow}  
\begin{algorithmic}[1]
    \Require graph structure  $G(V,E)$, vertex features $\{\textbf{h}^{0}_v\mid v\in V\}$, training set of vertices with ground-truth labels $V_{\mathbb{L}}$, number of layers $L$, initial model parameters $W^{l}_0$ for each layer $l$, and batch size $b$
  %number of layers $L$,  vertex labels $\{\mathbb{L}_v\mid v\in V_{\mathbb{L}}\}$, initial model parameters $W^{(l)}$ for each layer $l$, fanout $\{k_1, k_2, \cdots, k_L\}$, batch size $b$, number of iterations in an epoch $I$
    \Ensure updated parameters
    \State $\{V_0,\ldots V_I\} \leftarrow \texttt{split}{V_{\mathbb{L}}}$ by batch size $b$; \textcolor{gray}{//split training vertices into $I$ batches}
    %\State randomize vertex order in $V_{\mathbb{L}}$;
    \For{$i$ = $1$ to $I$}
        \State \textcolor{gray}{/* Sample a subgraph for a batch of training vertices */}
        \State $G_i\leftarrow \emptyset$; $V^F$= $V_i$; 
        \For{$l$ = $L$ to $1$} 
            \State $G^{l-1}(V^{l-1}, E^{l-1})\leftarrow$ \texttt{sample}($V^F$) \textcolor{gray}{// Sample one hop subgraph }
            \State $G_i=G_i\cup G^{l-1}$ ; $V^F\leftarrow V^{l-1}$
        \EndFor
            % \For{each $u$ in $V^l$}
            %         \State $V_{smpl}\leftarrow$ randomly select at most $k_l$ in-neighbors of $u$;
            %         \State $E_{smpl} \leftarrow$ the edges between $u$ and $V_{smpl}$;
            %         \State $V^{l-1} \leftarrow V^{l-1} \cup V_{smpl}$; $E^{l-1} \leftarrow E^{l-1}\cup E_{smpl}$;
            %     \EndFor
            %     \State $G_{i}\leftarrow G_{i}\cup V^{l-1} \cup E^{l-1}$;
            % \EndFor
        \State \textcolor{gray}{/* Gather the vertex features of sampled subgraph $G_i$ */}
        \State $H_i\leftarrow \{\textbf{h}_v^0\mid v\in G_i\}$
        \State \textcolor{gray}{/* Train $V_i$ with sampled subgraph $G_i$ and update parameters */}
        \For{$l$ = $1$ to $L$}
            \State Forward compute the vertex embeddings with $G_i, H_i, W^l_{i-1}$ based on Equation (\ref{form:agg_forward}) and (\ref{form:update_forward}); 
        \EndFor
        \State Compute the loss $\mathcal{L}$ and generate backward gradients;
        \For{$l$ = $L$ to $1$}
                \State Backward compute the gradients with $G_i, H_i, W^l_{i-1}$; 
                % based on Equation (\ref{form:agg_backward}) and (\ref{form:emb_backward});
                \State Update $W^{l}_i$ based on $W^{l}_{i-1}$ and $\nabla W^{l}_{i-1}$;
        \EndFor
    \EndFor                              

	\end{algorithmic}  
\end{algorithm}

\vspace{-0.3cm}
\subsection{Sample-based Mini-batch GNN Training}

Sample-based mini-batch GNN training splits the training vertices into multiple mini-batches. Each time it loads a batch of training vertices (with ground-truth labels) along with the sampled subgraph targeted at these vertices into GPUs for training \cite{Graphsage_2017,GCNWS_SIGKDD_2018,DLG_TKDE_2022}. Algorithm \ref{alg:sampleflow} elaborates the sample-based mini-batch GNN training process for a single epoch. It first splits training vertices $V_{\mathbb{L}}$ into $I$ mini batches according to the batch size $b$ (Line 1), then samples the multi-hop subgraph $G_i$ of each batch (Line 3-9). The sampled subgraph is generated by a reverse traversal from the training vertices (Line 5-7). There is a \textit{fanout} parameter to specify the number of sampled neighbors in the \texttt{sample} function per layer or per vertex for controlling the size of the sampled subgraph. It is noticeable that there exist other sampling schemes that help improve the accuracy, such as importance-aware sampling \cite{FASTGcn_iclr_2018}. We show a simple and general case since the sampling scheme is not our focus in this paper. According to the sampled subgraph $G_i$, we launch the gathering step to collect the features $H_i$ of sampled vertices (Line 9). The batch of training vertices $V_i$, the gathered features $H_i$, and sampled subgraph $G_i$ are then fed into the GNN model for training (Line 11-16). Specifically, there is a forward computation process of vertex embeddings (Line 11-12), loss computation with predicted labels and ground-truth labels (Line 13), and a backward computation process of gradients (Line 14-15). The layer-wise parameters $W^l$ are updated after each batch processing, so the parameters are updated multiple times instead of once in full-graph training.

%The algorithm first randomly splits batches of the training label vertices (target vertices) according to the batch size $b$ (Line 2). The number of iterations $I$ is the number of batches in an epoch, which represents that the model weights will be updated $I$ times. Compared with the full-graph training method, the model weights of the sample-based GNN training are updated more frequently. Then, to obtain sampled subgraph, the same graph sampling algorithms are utilized for each training vertex in a batch to select the random neighbors within $L$ hops (Line 3-8). The number of neighbors sampled in each layer depends on the fanout $\{k_1, k_2, \cdots, k_L\}$. Subsequently, the features of each vertex in the sampled subgraph are gathered into $requiredfeature$ and fed into the GNN model with $sampledgraph$ for training (Line 9-11). Finally, GNN training is performed on the mini-batched and sampled subgraph and their corresponding features (Line 12-16).
% For large graphs, mini-batch training can achieve faster convergence and higher accuracy \cite{distdglv2_sigkdd22}.

% \zyf{how is this subgraph generated? it seems that you first time mention subgraph but I don't know what subgraph is. I think the description is not clear. mini-batch (parameters updated more frequently than whole-batch), sample-based that generates subgraph. what do you mean by target vertex in fig 3. Is it possible to give a formal definition of the sampled subgraph? or an algorithm to describe the sample-based mini-batch training?}.

Figure \ref{fig:Sample} shows an illustrative example of sample-based mini-batch GNN training model mapping to CPU-GPU heterogeneous execution environment. This is a 2-layer GNN model on a graph of 9 vertices with a training set containing only a single vertex 4. 
The input data, including graph data and high-dimensional features, are typically stored in host memory, while GPUs perform the GNN training on mini-batched and sampled subgraphs.
In each training batch, this approach follows a \textbf{sample-gather-train} processing flow.
Firstly, the graph sampling algorithm uniformly selects two neighboring vertices for each vertex.
Secondly, the required vertex features are gathered based on the sampled subgraph and the training vertices. Then, they are loaded from the CPU to the GPU through a PCIe interconnect.
Finally, the GPU performs forward and backward computations, computing and updating the model parameters correspondingly.
\vspace{-0.05in}
\section{Existing systems and Their Limitations}
% \zyf{see comments}
% % frameworks是不是不太好？太大了？
\label{sec3}
%\wqg{这里你应该先说}

A key challenge in achieving high-performance sample-based GNN training is orchestrating tasks on the GPU-CPU heterogeneous environments.
% In Figure \ref{fig:system_classification}, we classify task orchestrating methods and list representative systems for each method. Among the eight possible methods, CPU-based training is not utilized if a GPU is available on the platform.
Existing sample-based GNN systems \cite{PAGRAPH_SOCC_2020,lazygcn_2020_neuips,Marius++_arxiv_2022,GNNlab_EUROSYS_2022,DATATIRING_VLDB_2022,PYTDIRECT_VLDB_2021,TurboGNN_TOC_2023,DGL_CORR_2019,wholegraph_SC_2022} typically separate the training process based on the operations, i.e., \textbf{sample}, \textbf{gather}, and \textbf{train}, and assign each operation to the GPU or the CPU. 
However, they often exhibit suboptimal performance due to inefficient CPU processing or GPU resource contention. To illustrate this, we conduct a series of experiments to analyze the problems of existing task orchestration methods.

\begin{figure*}
% \vspace{-0.3in}
  \centering
  \includegraphics[width=0.95\textwidth,page={1}]{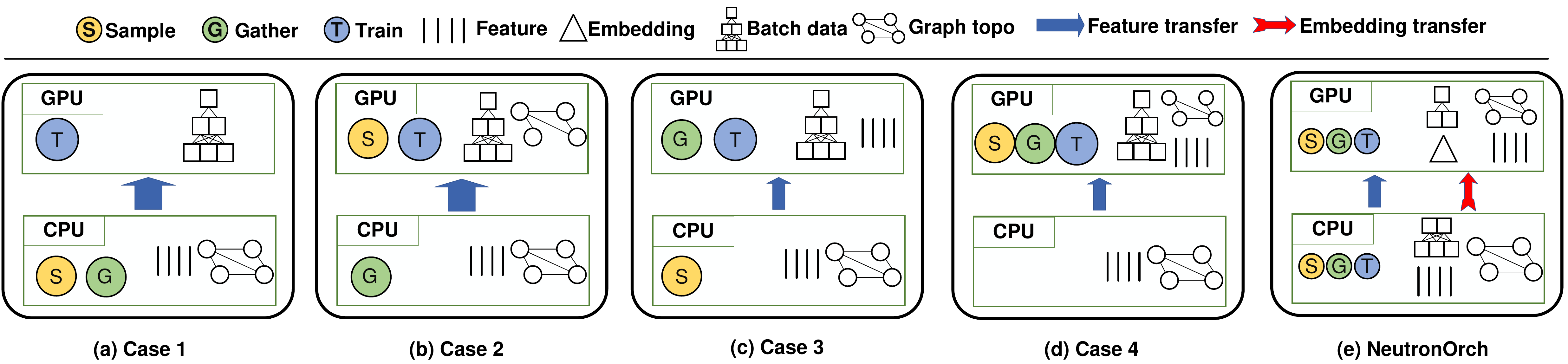}
  \vspace{-0.4cm}
  \caption{Illustration of the four task orchestrating methods and \system.
  (a) Case 1 \cite{lazygcn_2020_neuips,Marius++_arxiv_2022}: Placing sampling and gathering on CPUs. (b) Case 2 \cite{DGL_CORR_2019}: Placing sampling on the GPUs. (c) Case 3 \cite{PAGRAPH_SOCC_2020,refresh_arxiv_2023}: Placing gathering on the GPUs. (d) Case 4 \cite{DATATIRING_VLDB_2022,GNNlab_EUROSYS_2022,wholegraph_SC_2022,TurboGNN_TOC_2023,quiver_2023_arxiv,DSP_PPOPP_2023,ducati_sigmod_2023}: Placing sampling and gathering on the GPUs. (e) Hotness-aware layer-based task orchestrating method of \system. The width of the arrow is positively proportional to the transfer data volume. The volumes of the circles S, G, and T are proportional to the task amount.} 
  % \ccy{(Question: 1. Why is the feature represented by a vertical line instead of being represented by a rectangle consistent with Figure 3?  2. Blue arrows indicate transfer, written as "Feature". In addition, do you need to explain that the width indicates the transfer volume?)}
   \vspace{-0.2cm}
  \label{fig:task-method}
\end{figure*}

\vspace{-0.05in}
\subsection{Existing Task Orchestrating Methods}

In Figure \ref{fig:system_classification}, we classify task orchestrating methods and list representative systems. Among eight possible methods, CPU-based training is not utilized if a GPU is available on the platform. Therefore, we only 
consider placing sampling and gathering on different devices in the analysis. Further information regarding the used graphs, test platforms, and system configurations can be found in Section \ref{sec5}.

\Paragraph{Case 1: Placing sampling and gathering on CPUs suffers from inefficient CPU processing} We show an example of this case in Figure \ref{fig:task-method} (a). In this case, inefficient sampling and feature gathering on the CPU block the training pipeline, making data preparation the main bottleneck for overall performance.

\begin{table}[t]
	\vspace{-0.03in}
        \centering
    \caption{The runtime breakdown (in seconds) of sample and gather steps on DGL with different datasets. The FC and FT represent the feature collection and feature transfer of the gather step. GNN: A 3-layer GCN \cite{GCN_iclr_2017}.}
    \vspace{-0.1in}
    \label{tab:feature collection}
    \footnotesize
	\centering
	{\renewcommand{\arraystretch}{1.2}
 %\renewcommand{\arraystretch}{1}
	%\resizebox{0.97\linewidth}{!}{
	\begin{tabular}{l c c c c}
		\hline
		\multirow{1}*{\textbf{Dataset}}  & 
        \multirow{1}*{\textbf{Sample}}  & 
        \multirow{1}*{\textbf{Gather (FC)}}  &
        \multirow{1}*{\textbf{Gather (FT)}}  &
        \multirow{1}*{\textbf{Total}} \\
		\hline
    	% {Reddit}  & {0.84/23\%} & {0.608/17\%} & {0.372/10\%} & {3.63}\\ 
    	% {Lj-large} & {4.95/15\%} & {13.46/42\%} & {9.34/29\%} & {31.57}\\  
	    % {Orkut} & {2.43/10\%} & {9.94/43\%} & {8.11/35\%} & {22.93}\\      
	    % {Wikipedia} & {10.14/12\%} & {34.03/42\%} & {25.13/31\%} & {80.93}\\     
     %    {OGB} & {3.11/28\%} & {4.04/36\%} & {2.25/20\%} & {10.93} \\  
        {Reddit}  & {2.7/11\%} & {9.1/38\%} & {6.0/25\%} & {23.7}\\ 
    	{Lj-large} & {128.8/14\%} & {384.4/41\%} & {252.5/27\%} & {935.3}\\  
	    {Orkut} & {78.8/10\%} & {384.3/48\%} & {249.1/31\%} & {813.3}\\      
	    {Wikipedia} & {209.4/12\%} & {651.8/40\%} & {570.9/33\%} & {1669.1}\\     
        {Products} & {9.9/37\%} & {7.2/27\%} & {4.1/15\%} & {26.8} \\
        {Papers100M} & {11.5/32\%} & {8.6/24\%} & {6.4/18\%} & {36.84} \\  
		\hline
	
		\hline
	\end{tabular}
	}
 \vspace{-0.4cm}
\end{table}

%we often think of sampling as the main task in the CPU, while the data loading relies on PCIe. However, when we look further into the data loading step, we find that slow external interconnects are not the biggest bottleneck limiting the data loading performance. GNNlab \cite{GNNlab_EUROSYS_2022} and Pagraph \cite{PAGRAPH_SOCC_2020} as well as early DGL (DGL-UVA) \cite{DGL_CORR_2019} use explicit copy to transfer vertex features.

Neighbor sampling traverses the graph to obtain a multi-hop sampled subgraph for each training vertex. It incurs massive computation and irregular memory access, which makes the limited computational resources of the CPU hard to accelerate \cite{Nextdoor_eurosys_2021}. 
On the other hand, deploying the gathering step on the CPU, including storing vertex features in the host memory and performing feature collection and transfer, also proves to be inefficient \cite{PAGRAPH_SOCC_2020}. This inefficiency arises from two main factors.
Firstly, the slow external PCIe interconnect makes the host-GPU feature transfer much slower than GPU memory access \cite{PYTDIRECT_VLDB_2021}. Secondly, when loading the features of neighbors, collecting and organizing the fragmented vertex feature into contiguous memory before communication (to leverage the CUDA memory copy engine, \texttt{cudaMemcpy}) also consumes substantial CPU resources due to massive random memory accesses.
%\zyf{is it about CPU memory operation? why do you mention CUDA memory copy?}
%and sequential feature writing \zyf{sequential write of ... why sequential write is slow?}
We run this task orchestrating method with a 3-layer GCN on all six real-world graphs and analyze the time distribution for sampling and gathering steps.
As shown in Table \ref{tab:feature collection}, sampling and gathering steps occupy 19.3\% and 61.2\% of the total runtime, respectively. 
Specifically, the most significant overhead is the feature collection in the gathering step, which accounts for 36.3\% of the total runtime.

On the other hand, the long sampling and gathering time also causes low GPU utilization. The training step in the GPU needs to wait for the slow CPU tasks. % shows the average GPU utilization of the GCN trained on the Reddit dataset.
We can observe from Figure \ref{fig:utilization} that merely around 25\% of the GPU computational resources are utilized when the CPU executes both sampling and gathering steps. Even if only separate sampling or gathering is performed on the CPU, the overall GPU utilization is still less than 65\%. 
\Paragraph{Case 2: Placing sampling on the GPU suffers from GPU resource contention} We show an example of this case in Figure \ref{fig:task-method} (b). In this case, the slow sampling can be accelerated through the massively parallel processing capability of GPUs. However, the sampling and training operations will compete for the GPU resources, leading to suboptimal overall performance.

\begin{figure}[t]  
  \centering
  \includegraphics[width=7cm,page={1}]{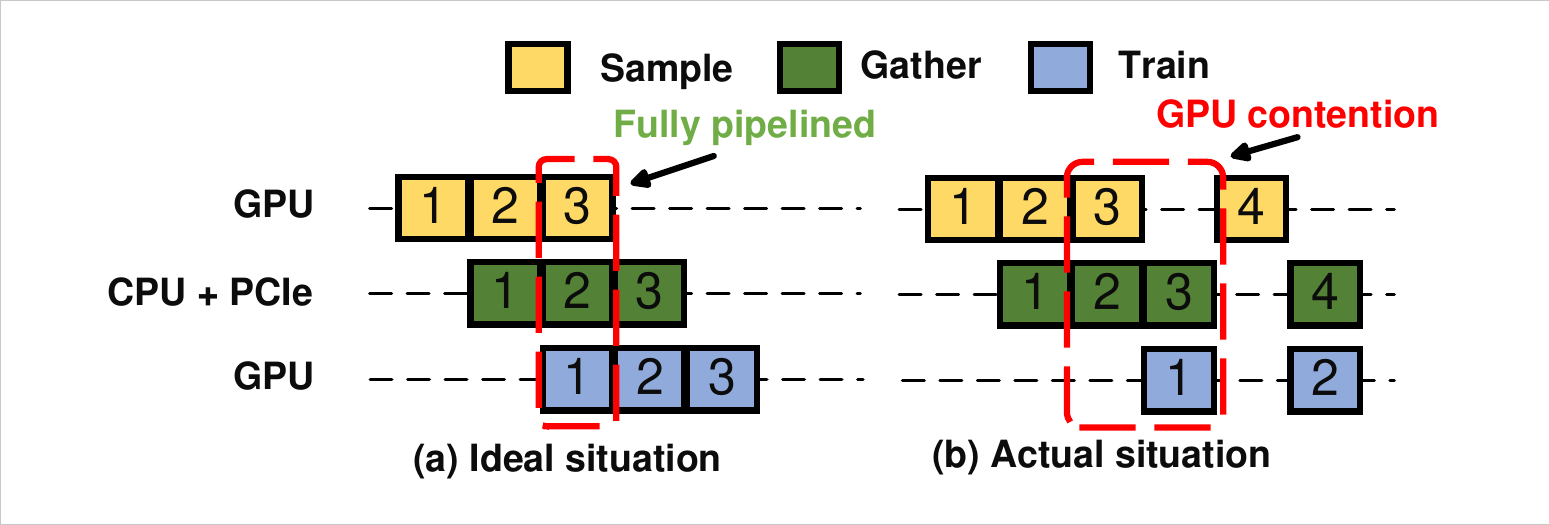}
  \vspace{-0.4cm} %调整图片与上文的垂直距离
  \caption{Two examples of multi-stream pipeline design in the ideal and actual situations.}
  \label{fig:pipeline}
   \vspace{-0.4cm}
\end{figure}

Sample-based GNN training typically employs a pipeline design that overlaps the different steps to achieve high performance in heterogeneous systems \cite{DGL_CORR_2019,PAGRAPH_SOCC_2020,GNNlab_EUROSYS_2022,Marius++_arxiv_2022}. Figure \ref{fig:pipeline} (a) depicts an ideal pipeline implementation where the operations of three batches (numbered 1, 2, and 3) are fully overlapped. However, this optimization requires different steps to be executed on independent resources. When placing sampling on the GPU, it competes for computation resources with the training step, leading to inefficient pipelining, as shown in Figure \ref{fig:pipeline} (b). To illustrate this, we evaluate the performance of DGL with pipelined optimization and non-pipelined under both CPU-based and GPU-based sampling. We run a 3-layer GCN on the Reddit graph and report the results in Table \ref{tab:breakdown-GPU_Sampling}. The GPU-based sampling with pipeline optimization reduces the runtime of its non-pipelined version by 43.1\%, which is a lower improvement than the effect of pipeline optimization on CPU-based sampling. Furthermore, with pipeline optimization, GPU-based sampling shows inferior overall performance compared to CPU-based sampling.

\begin{table}[t]
%\vspace{-0.03in}
    \setlength{\tabcolsep}{1mm}{
    \centering
    \caption{The runtime breakdown (in seconds) of a training epoch on DGL with CPU-based or GPU-based sampling. S, G, and T represent sample, gather, and train steps. GNN: A 3-Layer GCN \cite{GCN_iclr_2017}. Batch size 10000.}	
    \vspace{-0.1in}
    \label{tab:breakdown-GPU_Sampling}
	\centering
 \footnotesize
	{\renewcommand{\arraystretch}{1.2}
 %\renewcommand{\arraystretch}{1}
	%\resizebox{0.97\linewidth}{!}{
	\begin{tabular}{l c c c c c c c c}
		\hline
  
        \hline
		\multirow{1}*{\textbf{Configuration}}  &
        %\multirow{1}*{\textbf{GPU Utilization}} &
        \multirow{1}*{\textbf{S}} &
		\multirow{1}*{\textbf{G}}& {\textbf{T}} &{\textbf{Total}}&{\textbf{+pipeline}}\\
		\hline
    	% {CPU-based sampling}&{0.84}&{1.07} &{1.72} & {3.63}& {1.92} (-47.8\%)&\\ 
	    % {GPU-based sampling}& {0.33} &{0.93} &{1.74}& {3.08}& {2.03} (-34.0\%)\\
        {CPU-based sampling}&{2.28}&{2.84} &{2.76} & {7.88}& {3.42} (-56.6\%)&\\ 
	    {GPU-based sampling}& {0.78} &{2.69} &{2.75}& {6.22}& {3.54} (-43.1\%)\\  
	    %{GPU sampling + pipeline}  & {1.18} &{1.13} &{2.03} &{2.15} &\\     
		\hline
	
		\hline
	\end{tabular}}
	}
 \vspace{-0.3cm}
\end{table}

\begin{figure}[t]  
  \centering
  \includegraphics[width=0.45\textwidth,page={1}]{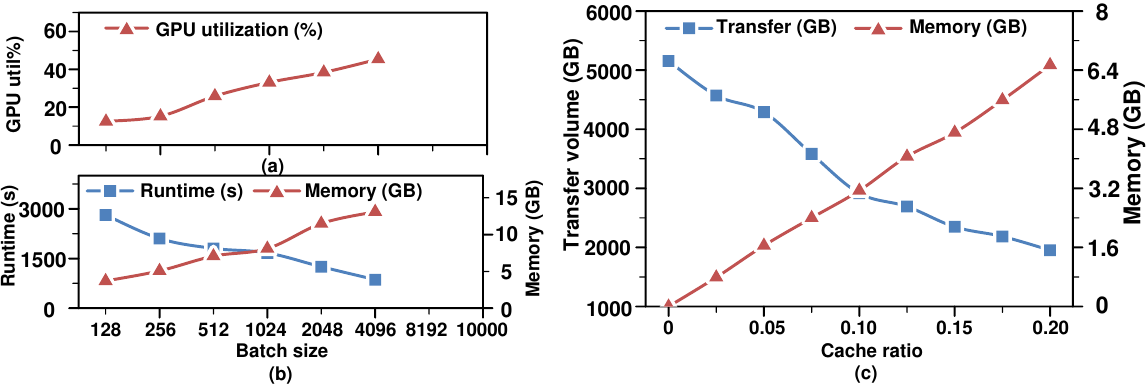}
    \vspace{-0.4cm}
   \caption{(a) GPU utilization with different batch sizes. (b) Per-epoch runtime and memory usage with different batch sizes. (c) Transfer volume and memory usage with different ratios of cached vertices (cache ratio).}
   \vspace{-0.4cm}
  \label{fig:batchsize-cache}
\end{figure}

\Paragraph{Case 3: Placing gathering on the GPU suffers from GPU memory contention} 
GPU-based gathering method leverages GPU memory to cache the vertex features, converting the slow host-GPU data communication into fast GPU memory access as much as possible \cite{PAGRAPH_SOCC_2020,GNNlab_EUROSYS_2022,DATATIRING_VLDB_2022,TurboGNN_TOC_2023,DSP_PPOPP_2023,ducati_sigmod_2023,legion_atc_2023,wholegraph_SC_2022}. However, the performance of GPU data caching is affected by the size of the available GPU memory. We show an example of this case in Figure \ref{fig:task-method} (c). In actual training, a substantial amount of global memory must be allocated for storing the training data and intermediate results, leaving only a small portion of memory available for feature caching. If using a large cache buffer, the memory allocated for training has to be reduced, resulting in the training with a small batch size. Unfortunately, this trade-off may lead to inferior performance since the computational power of GPUs cannot be fully utilized \cite{gnngaps_ppopp_2021,graphite_isca_2022}. 
To illustrate the impact of batch sizes and the ratio of cached vertices (denoted by cache ratio) on performance, we conduct experiments on DGL using the 3-layer GCN model on the Wikipedia graph. 
% \zyf{You always change GNN model and dataset, I suspect that you intend to gain the results you would like to see but this might not be true on other models/datasets.}
%去掉准确率实验，换为GPU利用率实验，准确率实验有争议，并不一定batchsize大就好，用wiki数据集
The results in Figure \ref{fig:batchsize-cache} (a) and (b) demonstrate that training with a larger batch size is beneficial for GNNs in terms of achieving better GPU utilization. Although this consumes more GPU memory, it results in faster execution.
The results in Figure \ref{fig:batchsize-cache} (c) reveal that a larger cache ratio results in a linear transfer reduction of features in the gathering step. 
However, when increasing the batch size from 128 to 4096, the ratio of cached vertices is decreased from 0.37 to 0.05 due to insufficient GPU memory, which results in numerous cache misses. Moreover, as the graph topology data and feature dimensions increase, the benefits brought by caching vertices will further diminish.

\Paragraph{Case 4: Placing sampling and gathering on the GPU suffers from GPU memory and resource contention} We show an example of this case in Figure \ref{fig:task-method} (d). When all steps of sample-based GNN training are executed on the GPU, it results in GPU contention and CPU idle. The reasons for this situation have been discussed in cases 2 and 3. Firstly, the computation kernels of sampling and training compete for the limited GPU cores, leading to a slowdown in both. Secondly, the batch data for training and cache data for gathering contend with the limited GPU memory. When further making the GPU responsible for the sampling step, the GPU needs to additionally hold the graph topology data, which can make the GPU memory contention worse.  
%As shown in Table \ref{tab:breakdown-SDT}, the GPU-led optimizations bring considerable acceleration to each step, and the pipeline is equally essential optimization because that can make full use of different resources in heterogeneous systems. Due to the GPU contention, when all optimizations are applied together, each step does not get the desired acceleration. Through a series of GPU-based optimizations, the existing systems can not only complete the sampling and training step in GPU but also use GPU-based caching policy to accelerate the data loading step. 

%Optimal end-to-end performance comes from pipeline optimization, which makes better use of heterogeneous resources. 

\vspace{-0.05in}
\subsection{Summary}
We conduct experimental analysis on different task orchestrating methods in GPU-CPU heterogeneous environments. 
%这里应该概要地总结一下他们有什么问题
Our observations reveal that step-based task orchestrating leads to an imbalanced allocation of computational and memory resources. Assigning two or more steps to the GPU may result in memory or GPU resource contention. On the other hand, assigning one or two steps to the CPU may cause inefficient CPU processing to become a bottleneck. 
%inferior performance
%which inevitably leads to resource contention at one end of the heterogeneous resource. After analysis and experiments, we summarize the major problems of the step-based computation orchestrating strategies into two types: (1) GPU resource contention. (2) Inefficient CPU processing. 
%Most applications developed for heterogeneous systems are more suitable for the GPU than the CPU. Only considering the device's suitability when scheduling the applications overloads the GPU and leaves the CPU idle. This results in a longer execution time. On the other hand, overusing the CPU and making the GPU wait only makes the situation worse. 
A well-designed CPU-GPU heterogeneous system should ensure adequate and balanced CPU and GPU utilization to achieve optimal performance. However, the step-based task orchestrating methods fail to achieve this. This motivates us to design a resource-balanced task orchestrating method.
\section{\system}
\label{sec4}

% \begin{figure}
%   \centering
%   \includegraphics[width=8cm,page={1}]{figures/Architecture.pdf}
%   \caption{Overall architecture of \system}
%    \vspace{-0.4cm}
%   \label{fig:architecture}
% \end{figure}

%改名 
%Motivated by the analysis and experimental results presented in Section \ref{sec2},
%Specifically, our architecture can achieve an average $\sim$ 90\% GPU utilization. 

We propose \system, a sample-based GNN training system that effectively improves CPU and GPU resource utilization through two critical techniques. 

\Paragraph{Hotness-aware layer-based task Orchestrating. }
Unlike step-based task orchestrating methods, \system decouples the training process by layers rather than steps and employs the sample-gather-train computation of each sub-task to a single device, eliminating the constraint of computing each step entirely on the CPU or GPU. To prevent CPU computation from becoming a bottleneck, \system allows the CPU to compute embeddings only for the hot vertices that are frequently accessed. Moreover, \system extends stale synchronous processing \cite{ssp_2013_nips} to guarantee bounded staleness of embedding reuse, thereby guaranteeing final accuracy.

\Paragraph{Super-batch pipelined training. } 
Concurrently executing sub-tasks deployed on the GPU and CPU is essential to achieve high performance. However, it is challenging to design a task scheduling method for layer-based task orchestrating because of the cross-layer data dependencies between the CPU and GPU training process. To solve this, we propose a super-batch pipelined training. In each super-batch, the CPU and GPU execute sub-tasks concurrently and independently of each other so that the training process is fully pipelined. In addition, this pipeline design ensures that the historical embeddings from the previous super-batch are only accessible within the current super-batch, naturally and efficiently enabling strict bounded staleness. 

\subsection{Hotness-Aware Layer-based Task Orchestrating}
In this section, we give a detailed discussion on the hotness-aware layer-based task orchestrating. It maximizes CPU-GPU resource utilization based on two principles. Firstly, partitioning the tasks in a fine-grained manner to balance the workloads on the CPUs and GPUs. Secondly, combining CPU and GPU resources to minimize CPU-GPU communication overhead. 

% Secondly, the computation in the CPU should be minimal yet valuable \zyf{minimal yet valuable, why not fully utilize CPU?}, enabling the CPU to assist the GPU in sharing computation tasks without making GPU wait.

%Firstly, the computation in the GPU should be adequate and reasonable, ensuring that GPU computing resources are fully utilized but do not cause memory or GPU resource contention. Second, the computation in the CPU should be minor and valuable, ensuring that the CPU can help the GPU share computation tasks and not cause the GPU to wait.

% \begin{table}%[!htbp]
% \vspace{-0.1in}
% \small
% \renewcommand{\arraystretch}{1.2}
% \centering
% 	\caption{Memory size of the feature and embedding for different hops in . Batch size :10000 Fanout: 4-4.}
% 	\label{tab:Memory size comparision}
% \begin{tabular}{|c|c|c|c|c|} 
% \hline
% Dataset & Hop & Vertex & Dimension & Memory (GB)\\ 
% \hline
% \multirow{2}*{Reddit}&source & 28706 &128 & 14.1\\ %600 * num * 4 / 1024 / 1024
% \cline{2-5}
% &destination & 66175 & 602 & 64.6\\

% \hline
% \multirow{2}*{Lj-large}& source & 24012 &128 &11.7\\
% \cline{2-5}
% &destination & 46475& 600 & 45.4\\

% \hline
% \multirow{2}*{Orkut}&source &24899 &128 &12.2\\
% \cline{2-5}
% &destination & 45813 &600 &44.7\\

% \hline
% \multirow{2}*{Wikipedia}&source &17216 &128 &8.4\\
% \cline{2-5}
% &destination & 26310 & 600 & 25.7\\

% \hline

% \end{tabular}
% \end{table}

\begin{figure}
  \centering
% \vspace{-0.05in}  
\includegraphics[width=8cm,page={1}]{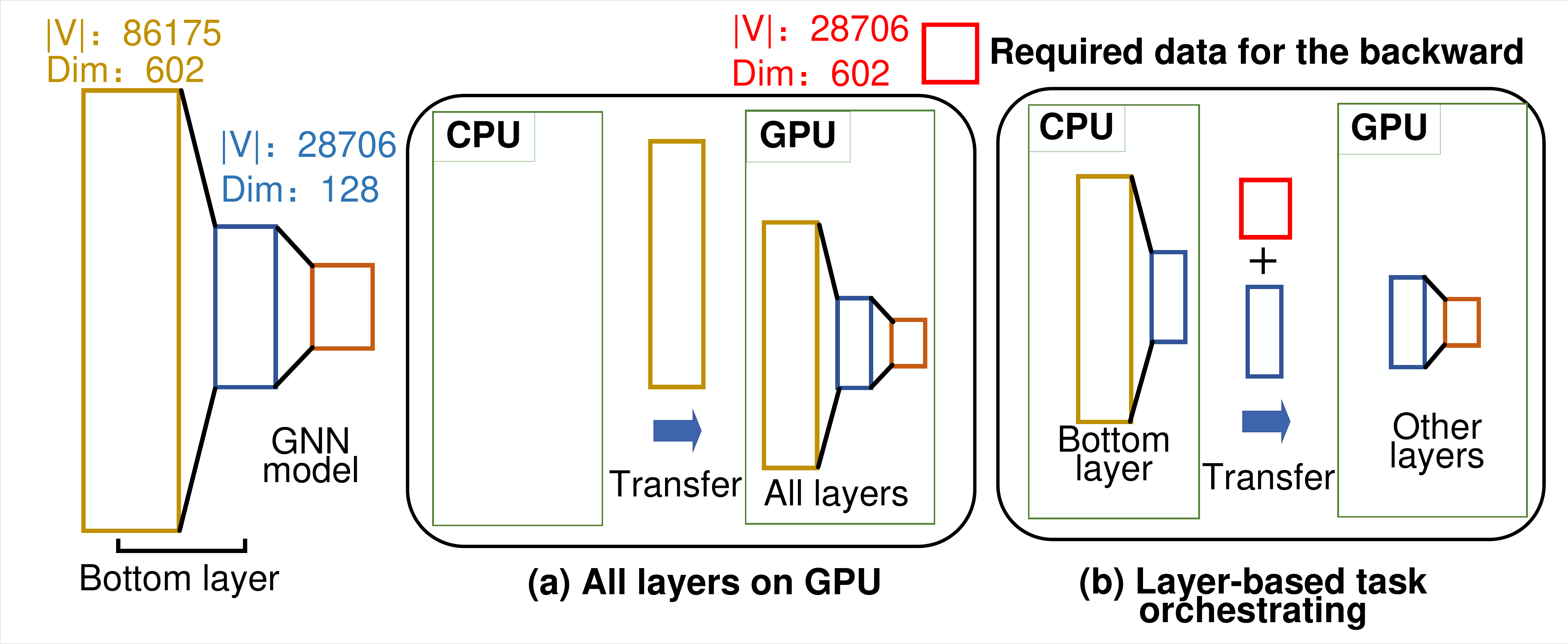}
     \vspace{-0.4cm}
      \caption{Workload and transfer volume of all layers on GPU and the layer-based task orchestrating, where |V| represents the number of vertices and Dim represents the dimension of features or embeddings. GNN: A 2-Layer GCN \cite{GCN_iclr_2017}. Dataset: Reddit.}
   \vspace{-0.3cm}
  \label{fig:layer-task}
\end{figure}

% Existing distributed GNN systems employ the data parallel training approach \cite{neutron_sigmod_2022,ROC_mlsys_2020,distgnn_sc_2021}. 
% These systems enable each vertex to collect its neighbors' embeddings from remote workers through communication. The graph propagation (including both the forward and backward propagation) occurs across workers. This design can be implemented by treating the CPU and GPU as separate workers and dividing the training process by layer between them. \zyf{I am not sure whether you want to mention distributed training. it might be unnecessary.} 

\subsubsection{Layer-based Task Orchestrating}
%Following these two principles, we decouple and analyze the sample-based GNN training to determine the task orchestrating method.
%We find that the training process of sample-based GNN training can be divided not only by the three steps but also by GNN layers. Fortunately, the CPU has a natural execution advantage for the bottom layer.

We find that the training process of sample-based GNN training can be divided not only by step but also by the GNN layer. As shown in Figure \ref{fig:task-method} (e), the CPU is responsible for the bottom layer computation of GNN training, while the GPU is responsible for the computation of the other layers. 
The CPU exhibits an inherent advantage when executing the bottom layer. Firstly, as the storage of features, the CPU can directly perform the GNN computation of the bottom layer without executing the time-consuming feature collection stage.
Secondly, the CPU-GPU communication overhead will decrease as the transfer objects are changed from the neighbors' features to the computed embeddings.
%The CPU can directly aggregate fragmented vertex features and write embeddings to continuous space after aggregation and update.
% \zyf{why not need feature collection stage?}
%of sampled vertices are aggregated and updated before the transfer begins.
% \wqg{comments: Lack of complete introduction or picture on Layer-based TO}
% %我咋感觉，你还没说 layer-based orchestrating 是啥呢？ 
% comment sanzo: backward pass or backward computation?
In layer-based task orchestrating, an additional data transfer stage is required in the backward pass for transferring the gradient back to the bottom layer training in the CPU. To improve training efficiency, we move the backward pass of the bottom layer to the GPU. Specifically, the required data for the gradient computation is transferred to the GPU along with the computed embeddings in the forward computation. Take the GCN \cite{GCN_iclr_2017} algorithm as an example. The aggregated neighbor representation and newly computed vertex embedding are transferred to the GPU.  

%毫无用处，且不知道在干什么的说法
%{we combine the two data transfers into one host-to-GPU data transfer.} 

%The required data for the backward pass is transferred together with the embeddings. 
%\zyf{the writing needs polish}  
% Compared to the traditional method, our method still decreases the total transfer volume because the aggregated results reduce the amount of the vertices to be transferred. 
% \zyf{too detailed, not reflected in the figure, the figure is hard to understand. Maybe you can use a concrete graph with 6-8 nodes associated with dimensional features to illustrate.}
%and four random neighbors are sampled for a vertex in each layer

We execute a two-layer GCN on the Reddit dataset to evaluate the impact of the layer-based task orchestrating method on the workload and transfer volume in the training process.
The training batch size is 10000, and four random neighbors are sampled for a vertex in each layer. We count the number of vertices and the feature or embedding dimension of each model layer.
As shown in Figure \ref{fig:layer-task}, the number of vertices in the bottom layer is three times that in the middle layer (86175 vs. 28706), positively related to the number of sampled neighbors (i.e., fanout). 
Meanwhile, the embedding dimension of the middle layer is generally smaller than the features in the outermost layer. Thus, despite introducing additional transmission for backward computation on the GPU, the layer-based task orchestrating method still has a smaller transfer volume.
On the other hand, this method makes more GPU memory available for training data, as it maintains the training data in both CPU and GPUs, resulting in reduced GPU memory usage. 

By employing layer-based task orchestrating, \system significantly reduces the computational workload and memory footprint of GPU training. 

\begin{figure}
  \centering
  \includegraphics[width=8cm,page={1}]{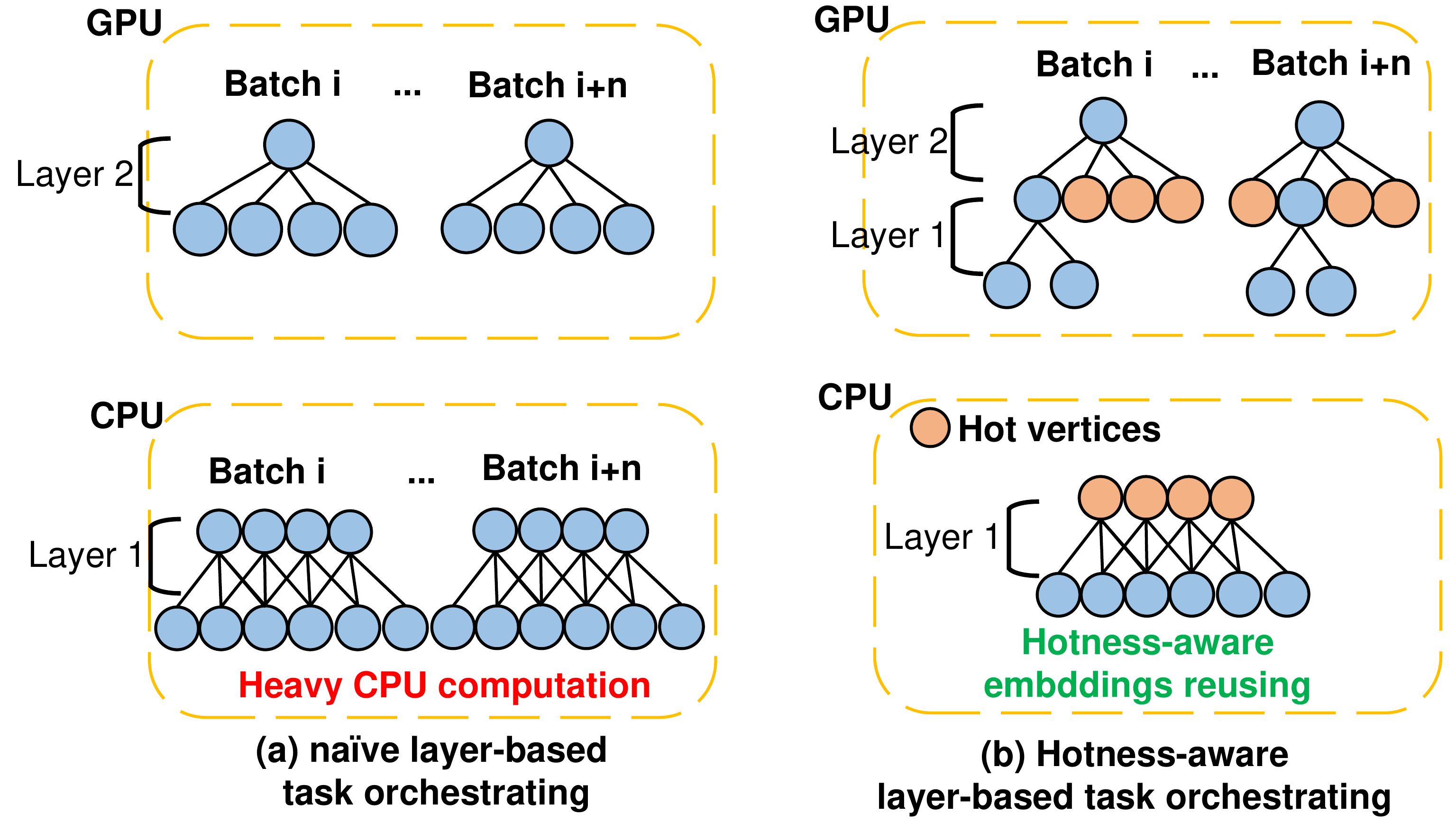}
     \vspace{-0.4cm}
  \caption{An illustrative of CPU and GPU workload in naive layer-based task orchestrating and hotness-aware layer-based task orchestrating.}
   \vspace{-0.2cm}
  \label{fig: computing framework}
\end{figure}

\subsubsection{Hotness-aware Embedding Reusing}
As shown in Figure \ref{fig: computing framework} (a), executing a complete bottom layer in the CPU may cause the CPU processing a new bottleneck because the computation volume of the entire layer is high.
% In other words, relying on the GPU as the primary computing power, we cannot afford to let it frequently wait for the CPU's results. This situation can lead to significant resource wastage.
Previous studies on caching policies have demonstrated that a subset of vertices is frequently accessed during sample-based GNN training \cite{PAGRAPH_SOCC_2020, DATATIRING_VLDB_2022, GNNlab_EUROSYS_2022}, which is commonly referred to as the "hot vertices". By caching the features of these vertices in the GPU, existing works can avoid repeatedly loading them from the host memory, thereby reducing CPU-GPU communications. 
% \zyf{logical gap, why caching can reduce transfer volume?}
\system follows similar ideas to design a hotness-aware computation offloading method, which computes only the embeddings for hot vertices in the CPU and allows the GPU to reuse these embeddings between batches within bounded staleness \cite{SANCUS_VLDB_2022, ST_ICML_2018,ssp_2013_nips}, our task orchestrating method can substantially decrease the CPU workload and CPU-GPU communication.

The main idea of reusing embeddings within bounded staleness is to maintain the historical embedding $\tilde{h}_i^{(l)}$ for exact embedding ${h}_i^{(l)}$ as an affordable approximation with a given bound \cite{ssp_2013_nips, ST_ICML_2018, SANCUS_VLDB_2022}. Bounded staleness expects $\tilde{h}_i^{(l)}$ and ${h}_i^{(l)}$ to be similar if the model weights do not change too fast during the training. In \system, we use the number of model parameter updates, i.e., the number of batches, as the bound. Specifically, if the bound is $N$, a recently computed embedding can be reused within the subsequent $N$ batches.
Using historical embeddings avoids the need for aggregating neighbor features and the associated backward pass.
This not only reduces the amount of raw features to load but also minimizes the CPU workload.
We will provide the theoretical analysis of the convergence guarantee for \system in Section \ref{prove}. 
Before the end-to-end training starts, \system requires a preprocessing stage to determine the hot vertices and set up a hot vertices queue for the processing. We employ the pre-sampling method of GNNLab \cite{GNNlab_EUROSYS_2022} to sample multi-hop neighbor multiple times for each training vertex and record the accessed frequencies (i.e., hotness) of the vertices. Then, we sort the vertices based on the hotness and determine the hot vertices used for CPU computation according to the hot vertex ratio.
 
Figure \ref{fig: computing framework} (b) illustrates an example of the hotness-aware layer-based task orchestrating method. During training, for hot vertices, the CPU samples them in one hop and computes their embeddings, which will be fetched and reused by the GPU training process to reduce computation and communication. For the cold vertices in the bottom layer, the GPU pulls the features of their vertices and computes them locally. 
To guarantee model accuracy, it is essential to control the version of reused embeddings within a specified range of batches during training, which is known as a concept of bounded staleness \cite{ssp_2013_nips}. To achieve this, we propose a super-batch pipelined training method that achieves bounded staleness among batches by packing each consecutive $N$ batches into a super-batch and controlling the embedding reusing among super-batches during computation. In this way, each embedding of the hot vertex is updated at least once within $N$ batches. A detailed discussion is provided in Section \ref{sec4.2}. Additionally, we offer theoretical analysis in Section 4.3 to demonstrate the correctness of our design. By adopting this approach, \system effectively offloads computations to the GPU and prevents CPU computation from becoming a performance bottleneck.

\subsubsection{Hybrid Hot Vertices Processing for high-performance GPU servers}
Orchestrating tasks across CPUs and GPUs balances CPU and GPU resource utilization and reduces CPU-GPU data communication by offloading the computation of hot vertices to the CPU. 
% However, in cases where the computing server is equipped with multiple GPUs but has limited CPU computing power, such as in a single-CPU-multi-GPU environment, the contribution of using CPU computation decreases because the improvement may not surpass the benefit brought about by the increased GPU resources. 
% 这里不是GPU计算算力的提高吧？感觉像是GPU内存的增加？或者GPU计算资源的充裕？（下面这句话用的是 GPU resources）
To address this problem, we propose a hybrid processing policy for hot vertices that further balances CPU and GPU utilization by assigning hot vertices to both CPU computation and GPU feature caching. When GPU resources significantly outnumber CPU resources, \system adaptively assigns hot vertices to the GPU, utilizing feature caching to avoid computing them on the CPUs. 
Specifically, during execution, \system monitors the time elapsed on GPU idleness caused by CPU computation and the remaining available GPU memory. 
It adjusts the allocation of hot vertices to GPU and CPU through a worklist, ensuring that GPU memory does not overly subscribe while minimizing GPU idle time. 
Note that \system stops assigning hot vertices to GPU when GPU memory is exhausted or the GPU idle time reaches zero, as appropriately utilizing CPU computation can effectively reduce data transfers. 
Compared to computing embedding of all hot vertices on the CPU, our hybrid processing optimization can further improve heterogeneous resource utilization.

\subsubsection{Discussion of \system and other HE-based GNN training frameworks.}
Recent studies \cite{ST_ICML_2018,SANCUS_VLDB_2022,GNNautoscale_icml_2021,refresh_arxiv_2023} have shown the benefits of using historical embedding (HE) for GNN training acceleration. VR-GCN \cite{ST_ICML_2018} uses HE in GNN training to reduce the number of vertices to sample. GNNAutoScale \cite{GNNautoscale_icml_2021} reuses HE for all vertices to reduce training memory consumption on the GPU.
Refresh \cite{refresh_arxiv_2023} caches HE in GPU memory to reduce the CPU-GPU feature communication and GPU training time. SANCUS \cite{SANCUS_VLDB_2022} reuses HE for boundary vertices to reduce communication in distributed full-graph training. 
However, these frameworks focus on scenarios where training is exclusively deployed on GPUs \cite{ST_ICML_2018,SANCUS_VLDB_2022,GNNautoscale_icml_2021,refresh_arxiv_2023}, and either target different training paradigms \cite{SANCUS_VLDB_2022}, uniformly reuse HE for all vertices \cite{ST_ICML_2018,refresh_arxiv_2023,GNNautoscale_icml_2021} which does not consider the impact of vertex hotness to the resource utilization, or do not strictly guarantee bounded staleness \cite{ST_ICML_2018, GNNautoscale_icml_2021} among batches, which is not applicable to \system that focuses on balancing GPU-CPU heterogeneous resource utilization and guaranteeing bounded accuracy. 
To achieve this goal, \system adopts several critical designs.
1) \system selectively computes HEs for the frequently accessed vertices and maximizes CPU-GPU resource utilization by adjusting the proportion of hot vertices assigned to CPU computation and GPU feature cache (Section 4.1.3).
2) We conduct a theoretical analysis on the convergence of \system's HE to demonstrate its correctness (Section 4.3). Furthermore, \system provides a super-batch pipelining method, combining HE version control with pipelined task scheduling to achieve efficient and strict bounded staleness (Section 4.2).

\begin{figure*}
  \centering
  \includegraphics[width=0.9\textwidth,page={1}]{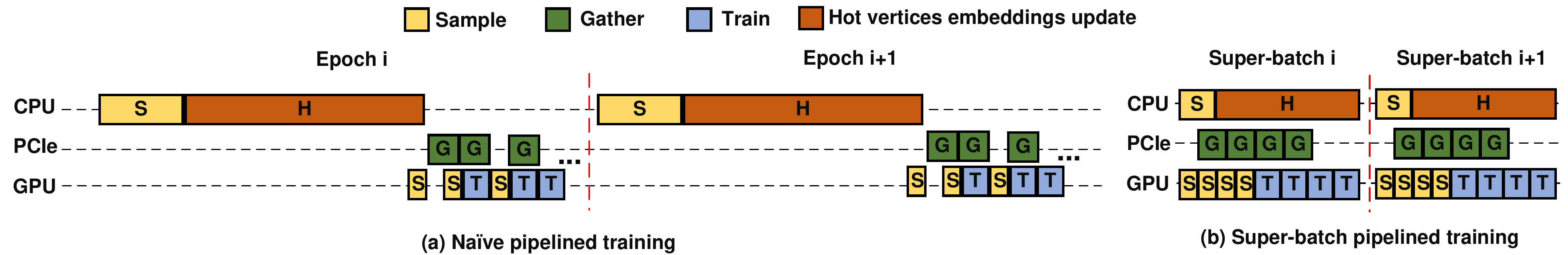}
     \vspace{-0.3cm}
  \caption{An illustrative example of super-batch pipelined training.}
   \vspace{-0.4cm}
  \label{fig:pipeline design}
\end{figure*}

% \zyf{I think the problem is not well-motivated. I don't understand what the problem is. Would it be better to formalize the problem?}

% We present a comprehensive exploration of the super-batch pipeline design, starting with an overview, followed by a detailed examination of each stage.

% \vspace{-0.05in}
\subsection{Super-batch Pipelined Training}
\label{sec4.2}

%我们讨论了由CPU和GPU交错计算所导致的数据依赖性，并提出。。。

In this section, we discuss the data dependencies arising from the integration of CPU and GPU computations and propose a super-batch pipelined training method to manage them efficiently. 
%感觉这句没啥问题啊

\subsubsection{Data Dependencies.} Seamlessly overlapping tasks across diverse computing resources through pipelining is essential to achieve high performance on heterogeneous systems \cite{PYTDIRECT_VLDB_2021,TurboGNN_TOC_2023,DSP_PPOPP_2023}. However, implementing high-performance task pipelining in \system is a non-trivial task because the layer-based task orchestrating creates cross-layer data dependencies between the CPU and GPU training process. 
Figure \ref{fig:pipeline design} (a) illustrates data dependencies in \system when setting the stale bound to one epoch \cite{ST_ICML_2018,GNNautoscale_icml_2021}.
The CPU updates the embeddings for hot vertices when it reaches the stale bound, and the training on the GPU must wait for the CPU to complete the embedding computation for hot vertices. Consequently, the CPU and GPU need to fall back to sequential computation, resulting in inefficient resource utilization.

%In addition, the model weights are updated after one epoch finishes. Setting one epoch as stale bound makes the embeddings for hot vertices too stale for batches that start training later. 

%Past works \cite{SANCUS_VLDB_2022,Marius_OSDI_2021,ssp_2013_nips} have shown that introducing asynchronous pipeline training and using stale embeddings within certain bound can speed up training and maintain accuracy.

% Therefore, selecting hot vertices from a limited number of upcoming batches is more reasonable and efficient than selecting hot vertices from the entire training set.
% \zyf{I don't know what the intuition is to merge n batches.} 
%In sample-based GNN training, each epoch contains many batches. As shown in Figure \ref{fig:pipeline design} (a), computing embeddings for hot vertices for all batches in an epoch is time-consuming and results in GPU resources idle.

\subsubsection{Pipeline Design.} 
To address the above issues, we propose a super-batch pipelined training method.
Firstly, the embeddings update for hot vertices in the CPU is partitioned into multiple sub-computing tasks, each of which computes the embeddings for hot vertices for subsequent multiple batches, which provides more opportunities for pipelined designs. Secondly, we limit the stale bound to several batches instead of one epoch because more tightly controlled bounded staleness can help ensure model accuracy.

Before training starts, \system combines $n$ batches into a single super-batch. Then, we select a hot vertices queue for each super-batch. As shown in Figure \ref{fig:pipeline design} (b), this is an illustrative example of a super-batch pipelined training when $n = 4$. During the pipelined training, the CPU computes embeddings for hot vertices for the next super-batch. In each super-batch, the GPU computes $n$ batches of training for a given batch size, sharing the historical embeddings provided by the CPU from the last super-batch.
%与每一次为一个epoch中所有热点更新嵌入相比，每次只为下一个super-batch更新嵌入的时间开销更小，同时减小了历史嵌入与真实嵌入之间的版本差距。
Compared to updating the embedding for all hot vertices, updating the embedding for only the next super-batch has less computation overhead and reduces the version gap between historical embeddings and exact embeddings.
% The super-batch pipeline has a shorter CPU embedding update period than selecting hot vertices from all training vertices, effectively maintaining bounded staleness.
Our pipeline design consists of four stages, each serving a specific purpose. We now elaborate on these individual stages.

% \zyf{may need rewrite, hard to follow.}
\Paragraph{Stage 1: Sampling.} 
The sampling tasks are divided between the CPU and GPU. Firstly, the CPU samples the hot vertices and generates a one-hop subgraph for bottom-layer training. Next, the GPU initiates n-hop sampling, and upon reaching the bottom layer, it skips the vertices previously sampled by the CPU. Meanwhile, the GPU completes $n$ rounds of sampling before $n$ training rounds to mitigate potential resource contention between the sampling and training kernels. 
%Moreover, CPU sampling is advanced at the start of the pipeline to make it easier for the GPU to benefit from the results. %The GPU will perform $K$ rounds of sampling, which will help the CPU finish training more hot vertices. 
% and incorporates the CPU's sampling results

\Paragraph{Stage 2: Embedding computation for hot vertices.} The CPU is responsible for updating the embeddings of the hot vertices intended for the subsequent super-batch within each super-batch. The CPU conducts training based on the hotness order of the vertices and utilizes the most recent version of parameters. 
% Once the CPU completes the sampling of hot vertices for the next super-batch, it promptly performs aggregation and updates on these vertices to obtain embeddings. 

%The CPU also keeps an array in pinned memory visible to both the CPU and GPU that records whether the hot vertices have completed bottom layer training and its version.
%In the backward, the CPU updates the parameters of the bottom layer based on the gradients transferred back from the GPU. As shown in Figure\ref{fig:pipeline design}, in order not to cause the GPU to wait for the backward phase of the CPU, the gradient will be transferred back at $k - 1$ batches and the model parameters in the bottom layer are updated once. 

\Paragraph{Stage 3: Feature gathering.} 
During the gathering step, the embeddings of hot vertices and the features of cold vertices are transferred. After being transferred, the embeddings of the hot vertices will be cached in the GPU for use in other batches within a super-batch. 
% The transfer object is determined by the CPU's sampling result and the hot vertices queue.
% The features corresponding to the non-hot vertices in the bottom layer are transferred based on their sampled vertices. 
%Additionally, the data necessary for the backward computation of hot vertices is transferred to the GPU along with the embeddings, ensuring version consistency.

%For the non-hot vertices in the bottom layer, the corresponding features are transferred according to their sampled vertices. The embeddings are transferred directly to the hot vertices, and cache the embeddings in the GPU for other batches in the super-batch. In addition, the data required for hot vertices backward is transferred to the GPU along with embedding, and the version is kept consistent. 
%In order to fully utilize PCIe and GPU resources, the GPU doesn't wait for the CPU to finish bottom layer training for all hot vertices. When the GPU ends sampling the first batch, it immediately starts data loading for the first batch. 
%If the hot vertices are sampled but not yet processed in the CPU, transfer the required feature to the GPU for training. Therefore, for $K$ rounds of data loading within a super-batch, the time required is gradually reduced as the embedding of the hot vertices is gradually cached in the CPU or GPU. 
%With this adaptive data loading design, the GPU benefits from the CPU's computing power without causing GPU to wait.
% \zyf{is these stages description too wordy? To me, it seems like no interesting idea is presented but only engineering details.}

\Paragraph{Stage 4: Training on GPU.}
The GPU training starts after $n$ rounds of GPU sampling. During the forward pass, the GPU pulls the historical embeddings of hot vertices in the bottom layer from either the CPU or the GPU cache. For other layers, the GPU performs aggregation and updates alternately as normal. In the backward pass, the GPU updates the parameters of each layer and shares the parameters of the bottom layer with the CPU. 

%For other layers, the GPU performs aggregation and updates alternately as normal. During the forward, the data required for backward of the hot vertices pulled from the CPU is also filled in the corresponding position. In the backward, the GPU updates the parameters of each layer and shares the parameters of the bottom layer with the CPU. 
%As shown in Figure \ref{fig:pipeline design} (b), this is an illustrative example of a super-batch pipelined training when $n = 4$.
%Therefore, the versions of the model parameters vary within a hot vertices queue. 
\Paragraph{Bounded Staleness.} 
Model parameters are updated once within each batch, so we record the model parameter version by batch number.
The CPU trains hot vertices in order of hotness and utilizes the latest model parameters. 
As shown in Figure \ref{fig:pipeline design} (b), during the first super-batch, the embeddings for hot vertices in the CPU may have versions ranging from $0$ to $n-1$, where $n$ is the number of batches contained within a super-batch. 
In the second super-batch, when the GPU pulls embeddings for hot vertices from the CPU, it may receive historical embeddings with model parameters that are older than the current version. 
The CPU must complete the embeddings update for hot vertices for the next super-batch within the current super-batch. This constraint ensures that the version gap remains within two super-batches, effectively preventing excessive staleness. Figure \ref{fig:pipeline design} (b) illustrates that the most significant version gap may occur in the last batch of the second super-batch. During this batch, the model parameter version is $2n - 1$, and it may utilize historical embeddings from the previous super-batch with a model parameter version of $0$. For the other batches, the version gap between historical embeddings and exact embeddings is smaller than the upper bound of $2n - 1$. 

\label{prove}
\subsection{Convergence Analysis}

%SANCUS performs full-batch GNN training without sampling, avoiding communication by caching historical embeddings with stale constraints. \system's hotness-aware design has a similar goal but is distinct from it in two aspects. Firstly, \system performs sampling-based GNN training in mini-batches. Secondly, \system only uses stale embeddings in the bottom layer. 
% To ensure high model accuracy, besides setting the bound for the version (batch number) bound of the stale embeddings, we also set the bound for the model parameters of the stale embeddings \cite{ST_ICML_2018}. In this process, we monitor the model parameter variation between adjacent super-batches, and the maximum model parameter variation $\epsilon{W}$ that can be tolerated be defined as $\begin{Vmatrix} \tilde{W} - W \end{Vmatrix}\le \epsilon{W}$, where $\tilde{W}$ is the model parameter of stale embeddings, $W$ is the model parameter of the current batch. If the model parameters changes between adjacent super-batches are within the bound, the GPU can use the stale historical embeddings from the last super-batch.
% We deduce the convergence guarantee of \system as follows.

In this section, we provide a theoretical analysis of convergence guarantee for \system.
Bounded staleness has been widely used by machine learning systems \cite{SANCUS_VLDB_2022,ssp_2013_nips,ST_ICML_2018,Marius_OSDI_2021,GNNlab_EUROSYS_2022}, and we present the theoretical analysis referring to the SANCUS \cite{SANCUS_VLDB_2022} and VR-GCN \cite{ST_ICML_2018}. 
To ensure bounded staleness, \system designs super-batch pipelined training to limit the version (batch number) bound to $2n$, where $n$ is the number of batches in a super-batch. In this process, we monitor the model weight variation between adjacent super-batches, and the maximum model weight variation $\epsilon{}$ that can be tolerated be defined as $\epsilon{} = max \bigtriangleup \begin{Vmatrix} W \end{Vmatrix} \times 2n$, where $max \bigtriangleup \begin{Vmatrix} W \end{Vmatrix}$ denotes the maximum value variation of $W$ in a model weight update. 
With this staleness bound, we deduce the convergence guarantee as follows.

\begin{itemize}[leftmargin=*]
    \item \textbf{Proposition 1} provides the necessary and fundamental inequality operations required for the theoretical analysis;
    
    \item \textbf{Lemma 1} states that by imposing bounded staleness on the weights, the approximations of the embeddings and intermediate matrix results are close to the exact results;
    
    \item \textbf{Lemma 2} further demonstrates that the approximations of gradients in the training process closely match the exact gradients;

    \item \textbf{Theorem 1} concludes that the weight changes during training occur at a sufficiently slow rate, ensuring that the gradients are asymptotically unbiased and guaranteeing convergence;
\end{itemize}

\textbf{Proposition 1.} \textit{Let} 
$\begin{Vmatrix} A \end{Vmatrix}_{\infty} = max_{ij} \begin{vmatrix}
A_{ij} \end{vmatrix}$, \textit{then we have}:
\begin{gather*}
\begin{Vmatrix} AB \end{Vmatrix}_{\infty} \le col(A) \begin{Vmatrix} A \end{Vmatrix}_{\infty} \begin{Vmatrix} B \end{Vmatrix}_{\infty} \\
\begin{Vmatrix} A \odot B \end{Vmatrix}_{\infty} \le \begin{Vmatrix} A \end{Vmatrix}_{\infty} \begin{Vmatrix} B \end{Vmatrix}_{\infty} \\
\begin{Vmatrix} A + B \end{Vmatrix}_{\infty} \le \begin{Vmatrix} A \end{Vmatrix}_{\infty} + \begin{Vmatrix} B \end{Vmatrix}_{\infty}
\end{gather*}
\noindent \textit{where}, $col(A)$ \textit{denotes the number of columns of the matrix} $A$, $\odot$ \textit{denotes the element wise product}.

This proposition has been proved by VR-GCN \cite{ST_ICML_2018}, and we omit the proof. We further denote $C$ as the maximum number of columns that exist in our analysis.

Lemma 1, lemma 2, and Theorem 1 have also been proved by SANCUS \cite{SANCUS_VLDB_2022} and VR-GCN \cite{ST_ICML_2018}, and we omit the proof and give the necessary and sufficient conditions for their establishment.

\textbf{lemma 1.} 
\textit{Assume all the activations are $\rho$-Lipschitz},  \textit{the} $\begin{Vmatrix} W_i \end{Vmatrix}_{\infty}$ and $\begin{Vmatrix} \hat{A}_i \end{Vmatrix}_{\infty}$ \textit{are bounded by some constant} $B$, \textit{and the historical weights} $\tilde{W}_i$ \textit{are close to the exact weights} $W_i$ \textit{with the staleness bound} $\epsilon$ \textit{where} $\begin{Vmatrix} \tilde{W}_i - W_i \end{Vmatrix}$ $\le$ $\epsilon$, $\forall i$. \textit{Then the approximation error of the stale embedding} $\tilde{H}$ \textit{and stale activation} $\tilde{Z}$ \textit{is bounded by some constant} $K$ \textit{that depends on} $\rho, C, B$: 
$\begin{Vmatrix} H_{i}^{l} - \tilde{H}_{i}^{l} \end{Vmatrix}_{\infty}$ $<$ $\epsilon K$, $\forall i$ $>$ $I$, $l = 1,..., L-1$;
$\begin{Vmatrix} Z_{i}^{l} - \tilde{Z}_{i}^{l} \end{Vmatrix}_{\infty}$ $<$ $\epsilon K$, $\forall i > I$, $l = 1,..., L$.

\textbf{lemma 2.}
\textit{Assume that activation function} $\sigma \left ( \cdot  \right )$ \textit{and the gradient} $\nabla \mathcal{L}$ \textit{are} $\rho-Lipschitz$, the $\begin{Vmatrix} \nabla \mathcal{L} \end{Vmatrix}_{\infty} $, 
$\begin{Vmatrix} \hat{A}_{i} \end{Vmatrix}_{\infty} $,
$\begin{Vmatrix} W_{i} \end{Vmatrix}_{\infty} $, \textit{and}
$\begin{Vmatrix} \sigma' \left ( Z_{i} \right ) \end{Vmatrix}_{\infty} $ \textit{are bounded by some constant} $B$, \textit{and the historical weights} $\tilde{W}_i$ \textit{are close to the exact weights} $W_i$ \textit{with the staleness bound} $\epsilon$ \textit{where} $\begin{Vmatrix} \tilde{W}_i - W_i \end{Vmatrix}$ $\le$ $\epsilon$, $\forall i$.
\textit{Then the approximation error of the gradient} $\tilde g(W_i)$ \textit{is bounded by some contents}: $ \begin{Vmatrix} \mathbb{E} \tilde g(W_i) - \nabla \mathcal{L}(W_i) \end{Vmatrix}_{\infty} \le \epsilon K$ \textit{and} $\forall i > I$, \textit{where K depends on} $\rho, C, B$.

%$\nabla_{\tilde{Z}^{(l)} }\tilde{\mathcal{L}}$ and
%, the initial weights $W_{(1)}$
\textbf{Theorem 1.} 
\textit{Given the local minimizer} $W^{\star}$.
\textit{Assume that 
(1) the activation} $\sigma(\cdot)$ \textit{is $\rho$-Lipschitz}, 
\textit{(2) the gradient of the loss function} $\nabla \mathcal{L}(W_i)$ \textit{is $\rho$-Lipschitz and bounded}, 
\textit{(3) The gradient matrices}
$\begin{Vmatrix} \tilde{g}(W) \end{Vmatrix}_{\infty}$, 
$\begin{Vmatrix} g(W) \end{Vmatrix}_{\infty}$ and 
$\begin{Vmatrix} \nabla \mathcal{L}(W) \end{Vmatrix}_{\infty}$ \textit{are bounded by some constant} $G > 0$. 
\textit{(4) The loss} $\mathcal{L}(W)$ \textit{is $\rho$-smooth}, \textit{i.e.},
$$\lvert \mathcal{L}(W_2) - \mathcal{L}(W_1)- 	\left \langle \nabla \mathcal{L}(W_1), W_2 - W_1  \right \rangle \rvert \leq \frac{\rho}{2} \begin{Vmatrix} W_2 - W_1 \end{Vmatrix}_{F}^{2}, \forall W_1, W_2$$
, \textit{where $\left \langle A, B \right \rangle = tr(A^{T}B)$ is the inner product of matrix $A$ and matrix $B$}. 
\textit{Then, there exists $K > 0$, s.t., $\forall N > I$, if we run SGD for $R \leq N$ iterations, where R is chosen uniformly from $[I+1, ..., N ]$ and the learning rate }
$\eta = min\{{\frac{1}{\rho}, \frac{1}{\sqrt{N}}\}}$, , we have: 
$$\mathbb{E}_{R} \Vert \nabla \mathcal{L} (W_{R}) \Vert_{F}^{2} \leq 2\frac{\mathcal{L} (W_{1})- \mathcal{L}(W^{\star}) + \frac{\rho K}{2}}{\sqrt{N}}$$

when $N \rightarrow \infty$ , $\mathbb{E}_{R} \Vert \nabla \mathcal{L} (W_{R}) \Vert_{F}^{2} \rightarrow 0$, the above concludes that the convergence is guaranteed.

\section{Evaluation}
\label{sec5}
% We implement \system based on NeutronStar \cite{neutron_sigmod_2022} and LibTorch (v1.9) \cite{pytorch_NeurIPS_2019}. 
This section evaluates the performance of \system using three representative GNN models and six real-world graph datasets. 
%\zyf{the following words can be removed.} Specifically, we investigate the overall performance and the efficiency achieved through hotness-aware layer-based task orchestrating and super-batch pipelined training. Additionally, we conduct a breakdown analysis under various conditions.

\subsection{Experimental Setup}

\Paragraph{Environments.}
%单机单卡
Single GPU experiments are conducted on an Aliyun server equipped with an Intel Xeon Platinum 8163 CPU (48 cores and 368 GB main memory) and NVIDIA V100 (16GB) GPU. The multi-GPU experiments are conducted on an Aliyun server equipped with an Intel Xeon Platinum 8163 CPU (96 cores and 736 GB main memory) and eight NVIDIA V100 (16GB) GPUs. The eight GPUs are connected to the CPU via four PCIe-3.0 switches and equipped with NVLink interconnects similar to NVIDIA DGX-1 \cite{DGX1}. The GPU is enabled with CUDA 11.4 runtime and 418.67 drivers, and the host side is running Ubuntu 18.04 with Linux kernel version 4.13.0. All the source codes are compiled with O3 optimization.

%We adopt the same model architectures and configurations as in the reference.
\Paragraph{GNNs and model configurations.}
% We evaluate \system using two representative GNN models: Graph Convolutional Network (GCN) \cite{GCN_iclr_2017} and GraphSAGE \cite{Graphsage_2017}.
We evaluate \system using three representative GNN models: GCN \cite{GCN_iclr_2017}, GraphSAGE \cite{Graphsage_2017}, and GAT \cite{GAT_ICLR_2018}. Compared with other systems, the training batch size is set to 1024, the model depth is set to 3, and the sampling fan-out is set to [25, 10, 5]. When using more layers for performance evaluation, sampling fan-out beyond 3 layers will be set to 5, i.e., [25, 10, 5, 5 $\cdots$].

% Graph Convolutional Network (GCN) utilizes 2-hop random neighborhood sampling and GraphSAGE utilizes 3-hop random neighborhood sampling. Graph Convolutional Network (GCN) \cite{GCN_iclr_2017} generalizes the graph convolution operation as follows. Each vertex in a GCN layer aggregates the features of its neighboring vertices using a summation operation. Subsequently, the aggregated feature is passed through a fully-connected layer followed by a ReLU activation to generate output representations. GraphSAGE \cite{Graphsage_2017} is an inductive learning model that employs different aggregation functions depending on the number of hops. GraphSAGE offers four aggregation types: GCN(sum), Mean, LSTM, and Pooling. We present the results of the GraphSAGE-Mean model because all four aggregators exhibit similar execution costs \cite{Graphsage_2017}. 
% %Past works \cite{sampleacc_icml_2018,ST_ICML_2018,GNNSampling_sigops_2021, sampling_caa_2022} have demonstrated that sampling-based training can achieve comparable model accuracy to full graph training, even when sampling only a few neighbors at each neighborhood layer.
% In our evaluation, we set the sampling fanout to 4 for every layer. In addition, we set the training batch size to 10000 as it typically allows for full utilization of GPU.

\begin{table}[!t]
\vspace{-0.05in}
	\caption{Dataset description.}
	\vspace{-0.1in}
	%\hspace{-0.5in}
	\label{tab:Dataset}
	\centering
	\footnotesize
	{\renewcommand{\arraystretch}{1.2}
	\begin{tabular}{l r r c c c c}
		\hline
		
		\hline
		{\textbf{Dataset}} &
		{\textbf{|V|}} &
		{\textbf{|E|}}  &
		{\textbf{ftr. dim}}&
		{\textbf{\#$\mathbb{L}$}}&
		{\textbf{hid. dim}}\\
		\hline
        {Reddit \cite{Graphsage_2017}} & 232.96K &114.61M & 602 &41 &256\\
		{Lj-large \cite{LIVE_KDD_2006}} & 10.69M & 224.61M & 400 & 60 &256\\	%没有feature
		{Orkut \cite{ORUKT_KIS_2015}} &3.1M& 117M &600&20&160\\	%没有feature
        {Wikipedia \cite{konect_www_2013}} & 13.6M &437.2M & 600 &16&128\\
        {Products (PR) \cite{ogb_2020_neurIPS}} & 2.4M &61.9M & 100 &47&64\\
        {Papers100M (PA) \cite{ogb_2020_neurIPS}} & 111M &1.6B & 128&172&64\\
		\hline
		
		\hline
		
	\end{tabular}
	}
	\vspace{-0.1in}
\end{table}

% {ogbn-products} & 2.4M &61.9M & 128 - 4096 &16 &256\\
% {amazon}& 14.7M  & 64.0M & 578 & 16&160\\
% {ogbn-papers100M} & 111.1M & 1.6B &128&41&256\\	
% \begin{table}[!t]
% \vspace{-0.05in}
% 	\caption{The competitor systems.} 
% 	\vspace{-0.1in}
% 	%\hspace{-0.5in}
% 	\label{tab:system}
% 	\centering
% 	\footnotesize
% 	{\renewcommand{\arraystretch}{1.2}
% 	\begin{tabular}{l c c c c}
% 		\hline
		
% 		\hline
% 		{\textbf{System}} &
% 		{\textbf{Sample}}  &
% 		{\textbf{Gather}}&
% 		{\textbf{Train}}\\
% 		\hline
% 		{DGL \cite{DGL_CORR_2019}} &CPU & CPU & GPU\\
% %		\hline
% 		{PaGraph \cite{PAGRAPH_SOCC_2020}} & CPU & GPU & GPU\\
% %		\hline
% 		{GNNLab \cite{GNNlab_EUROSYS_2022}} & GPU & GPU & GPU\\		
           
%         {DGL-UVA \cite{PYTDIRECT_VLDB_2021}} & GPU & CPU &GPU\\
        
%         {GAS \cite{GNNautoscale_icml_2021}} & GPU & CPU & GPU \\
        
%         % {DSP \cite{DSP_PPOPP_2023}} & GPU & GPU &GPU \\
        
%         {\system (ours)} & CPU-GPU & CPU-GPU &CPU-GPU\\	
% 		\hline
		
% 		\hline
		
% 	\end{tabular}
% 	}
% 	\vspace{-0.1in}
% \end{table}

\Paragraph{Datasets.} 
For evaluation, we utilize {six} real-world graph datasets, as listed in Table \ref{tab:Dataset}. These include Reddit \cite{Graphsage_2017} and Orkut \cite{ORUKT_KIS_2015}, which are social networks, the Wikipedia (Wiki) network \cite{konect_www_2013} comprising wikilinks from the English Wikipedia, the LiveJournal communication network (Lj-large) \cite{LIVE_KDD_2006}. The Products (PR) \cite{ogb_2020_neurIPS} dataset is based on Amazon's co-purchasing network. The Papers100M (PA) \cite{ogb_2020_neurIPS} is a citation graph, where each vertex represents a paper and the edge represents the citation relation.
The "ftr. dim" column represents the dimension of vertex features, the "\#L" column represents the number of vertex classes, and the "hid. dim" column represents the embedding dimension of the hidden layer output.
For graphs without ground-truth properties (Lj-large, Orkut, and Wikipedia), we use randomly generated features, labels, training (65\%), test (10\%), and validation (25\%) set division.

%that incorporates CPU-based graph sampling and CPU-based feature gathering
%that incorporates GPU-accelerated graph sampling and zero-copy optimizations

% In addition, we compare \system with GNNAutoScale (GAS) \cite{GNNautoscale_icml_2021}, a representative system based on historical embedding.

% PaGraph designs a degree-based feature cache policy to reduce CPU-GPU communication. GNNLab proposes a pre-sampling-based feature cache policy to accommodate different graph datasets and sampling algorithms. stores graph topology and popular vertex features among the GPUs, fully utilizing the memory of the GPUs and reducing the CPU-GPU communication prunes the computation graph and reduces the GPU memory requirement of GNN training by utilizing historical embeddings

\Paragraph{Baselines.}
We compare \system with DGL \cite{DGL_CORR_2019}, DGL-UVA \cite{PYTDIRECT_VLDB_2021} , PaGraph \cite{PAGRAPH_SOCC_2020}, GNNLab \cite{GNNlab_EUROSYS_2022}, GNNAutoScale (GAS) \cite{GNNautoscale_icml_2021}, and DSP \cite{DSP_PPOPP_2023}. 
All comparison systems use GPUs to conduct training.
DGL and DGL-UVA store both the graph structure and features in CPU memory. The difference is that DGL conducts sampling using the CPU, while DGL-UVA conducts sampling using the GPU by utilizing the UVA technique.  
PaGraph and GNNLab utilize GPU-based feature caching to reduce CPU-GPU communication. PaGraph conducts the sampling using the CPU, and GNNLab stores the graph structure in GPU memory and conducts the sampling using the GPU.
GAS conducts feature gathering in the CPU and utilizes historical embedding to accelerate training. 
% The utilization of historical embeddings in GAS is relatively coarse-grained. Initially, GAS calculates and saves historical embeddings for all vertices. However, GAS does not provide bounded staleness and historical embeddings are reused among an epoch. 
%GAS 对历史嵌入的利用相对粗粒度。最初，GAS 会计算并保存所有顶点的历史嵌入。但是，GAS 并不提供有界限的滞后性，历史嵌入会在一个时代内重复使用。
DSP is a multi-GPU GNN training system that uses multi-GPU cooperative sampling. It also caches the graph topology and popular vertex features in GPU memory to accelerate the gathering step.
All these systems use the four specific step-based task orchestrating methods as discussed in Section \ref{sec3}. 
In the single GPU experiments, we compare \system with five systems: DGL \cite{DGL_CORR_2019}, DGL-UVA \cite{PYTDIRECT_VLDB_2021} , PaGraph \cite{PAGRAPH_SOCC_2020}, GNNLab \cite{GNNlab_EUROSYS_2022}, and GAS \cite{GNNautoscale_icml_2021}.
As GAS only provides a single GPU implementation, in multi-GPU experiments, we compare \system with four systems: PaGraph \cite{PAGRAPH_SOCC_2020}, DGL-UVA \cite{PYTDIRECT_VLDB_2021}, GNNLab \cite{GNNlab_EUROSYS_2022}, and DSP \cite{DSP_PPOPP_2023}.
% We present the details of the compared systems in Table \ref{tab:system}.

%DGL, DGL-UVA, PabGraph, and GNNLab are typical GNN systems using the four specific step-based task orchestrating methods as discussed in Section \ref{sec3}. 

\begin{figure}
  \centering
\includegraphics[width=8cm,page={1}]{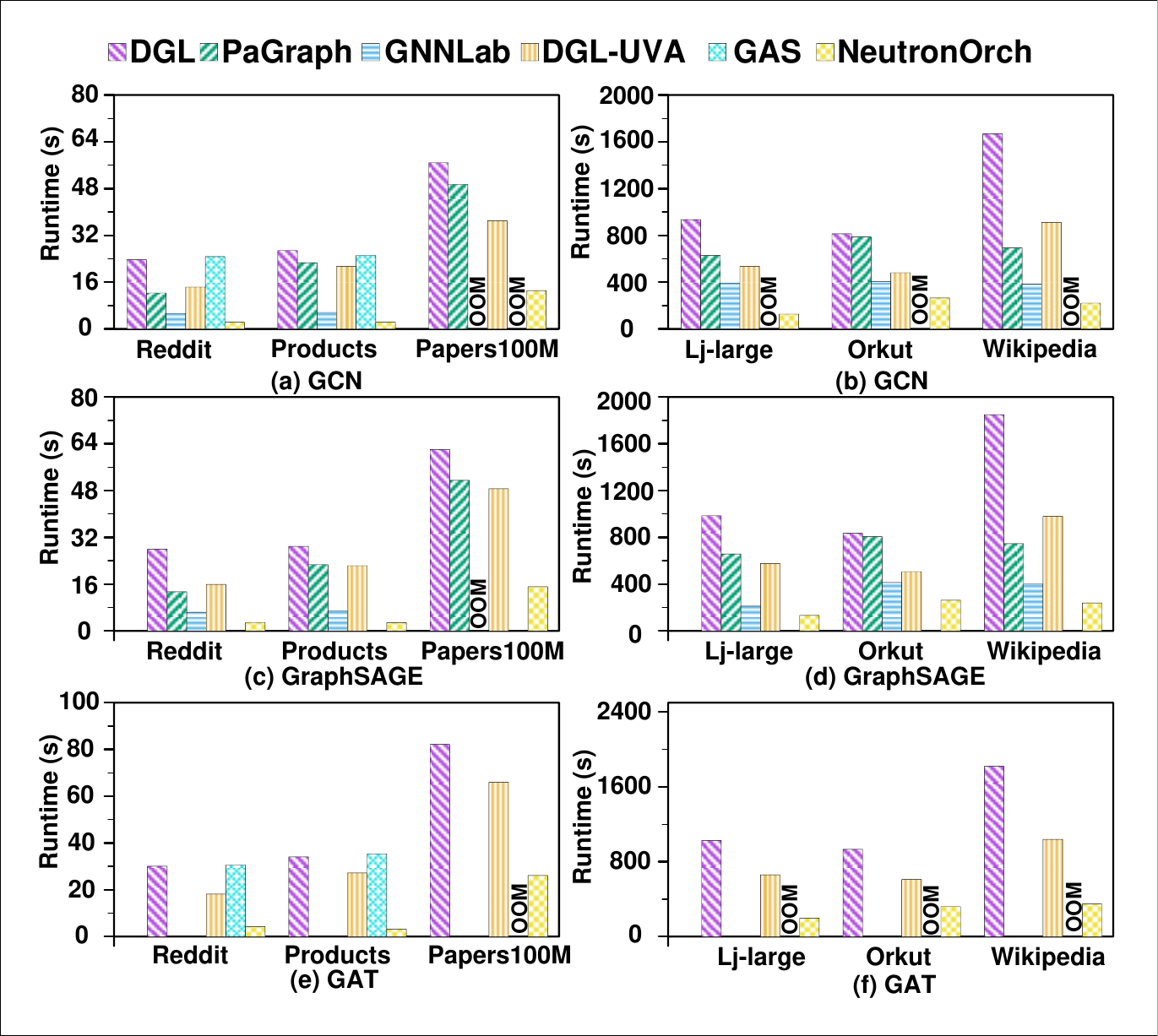}
    \vspace{-0.4cm}
  \caption{Overall training performance comparison (``OOM'' denotes out of memory).}
   \vspace{-0.6cm}
  \label{fig:overall-performance}
\end{figure}

\subsection{Single GPU Performance}
%We first compare \system with its competitors. 
Figure \ref{fig:overall-performance} shows the average training time of one epoch for all GNN systems. Note that GNNLab \cite{GNNlab_EUROSYS_2022} and PaGraph \cite{PAGRAPH_SOCC_2020} do not provide GAT support, and GAS \cite{GNNautoscale_icml_2021} does not provide GraphSAGE support.

(1) Comparison with DGL \cite{DGL_CORR_2019}: \system achieves a speed-up ranging from 2.91$\times$ (for GAT on Orkut) to 11.51$\times$ (for GCN on PR) compared to DGL.
DGL exhibits inferior performance compared to the other systems due to inefficient CPU-based sampling and gathering steps.
% As discussed in Section \ref{sec3}, DGL's task orchestrating method requires the CPU to perform both the sampling step and feature collection in the gathering step. 
The CPU resources are insufficient for effectively accelerating the sampling step in parallel and are also unable to handle the frequent random memory access required for feature collection. These two steps account for a substantial portion of the total runtime, resulting in considerable waiting time for the GPU.
\system effectively avoids inefficient CPU processing by selectively computing historical embeddings for hot vertices on the CPU. Hot vertices typically account for only a fraction of all training vertices, and historical embeddings can be reused multiple times across multiple batches.
% , effectively reducing computation on the CPU. 

%Second, the training data, mainly the features of the hot vertices' sampled neighbors, are stored in the CPU memory. 
(2) Comparison with PaGraph \cite{PAGRAPH_SOCC_2020}: \system achieves a speed-up ranging from 2.68$\times$ (for GraphSAGE on PA) to 9.72$\times$ (for GCN on PR) compared to PaGraph. PaGraph uses GPU-based gathering, reserving a portion of GPU memory for caching the features of high-degree vertices. The performance of PaGraph is limited by inefficient CPU sampling and GPU memory contention. 
\system provides more flexible task orchestration based on heterogeneous resources. When GPU resources are limited, increasing the CPU embedding computation reduces the memory requirement for training and caching in the GPU. When GPU resources are sufficient, reduce CPU embedding computation while increasing the feature cache ratio to avoid GPU waiting.
%As discussed in Section \ref{sec3}, the hit ratio of cache policy decreases as the feature dimension and batch size increase due to the limited GPU memory. 
% By employing the hotness-aware layer-based task orchestrating method, \system is memory-efficient. First, NeutronOrch saves the training memory requirement by offloading the computation to the CPU. Second, the cached objects of \system are only historical embeddings for hot vertices within a super-batch, consuming only a small amount of GPU memory. Therefore, \system can enable more GPU memory for accelerating sample and train steps while avoiding GPU memory contention. 

(3) Comparison with GNNLab \cite{GNNlab_EUROSYS_2022}: \system achieves a speed-up ranging from 1.52$\times$ (for GCN on Orkut) to 2.43$\times$ (for GraphSAGE on PR) compared to GNNLab.
NeutronOrch performs better when training with larger models or datasets. GNNLab needs to maintain the graph topology, cache data, and training data in the GPU memory, which constrains its scalability, especially when training billion-scale graphs like Papers100M for users with limited GPU resources. When the training memory requirement increases, GNNab's performance degrades due to a decreased cache hit rate. Moreover, when handling deeper GNN models, GNNLab encounters out-of-memory (OOM) issues due to GPU memory exhaustion, as shown in Table \ref{table:vary_layer}, Table \ref{table:vary-batchsize}, and Section 5.7.

%In contrast, NeutronOrch reduces computation, communication, and memory consumption by reusing historical embeddings and scheduling training tasks across CPUs and GPUs. As shown in Table \ref{table:vary_layer}, when training a 3-layer, 4-layer, and 5-layer GCN on Products, our speedups are 1.80$\times$, 1.89$\times$, and 2.24$\times$, respectively. 

%GNNLab proposes a factored system that places sampling and training steps across GPUs and introduces a pre-sampling-based caching policy to accelerate the data loading step. However, in a heterogeneous environment with only one GPU, as discussed in Section \ref{sec2}, the three steps executed on the GPU can lead to memory and resource contention issues. The GPU resource contention and CPU computing power idling can both adversely affect the performance of heterogeneous systems. Through the hotness-aware layer-based task orchestrating method, \system allocates reasonable and sufficient tasks for both CPU and GPU, avoiding resource contention. On the other hand, the super-batch pipeline design ensures that \system has high CPU and GPU utilization. 

%By assigning rational computational tasks to the CPU, \system fully utilizes memory and computation resources in heterogeneous environments.
% By capitalizing on increased CPU and GPU utilization simultaneously, \system demonstrates better overall performance.
% \wqg{What engine does DGL use?}
% \wqg{If it is deployed in \system, then what are the differences between our system and DGL-UVA}, which is also deployed on the \system

\begin{figure*}[th]
  \centering
  \vspace{-0.3in}
  \includegraphics[width=0.9\textwidth,page={1}]{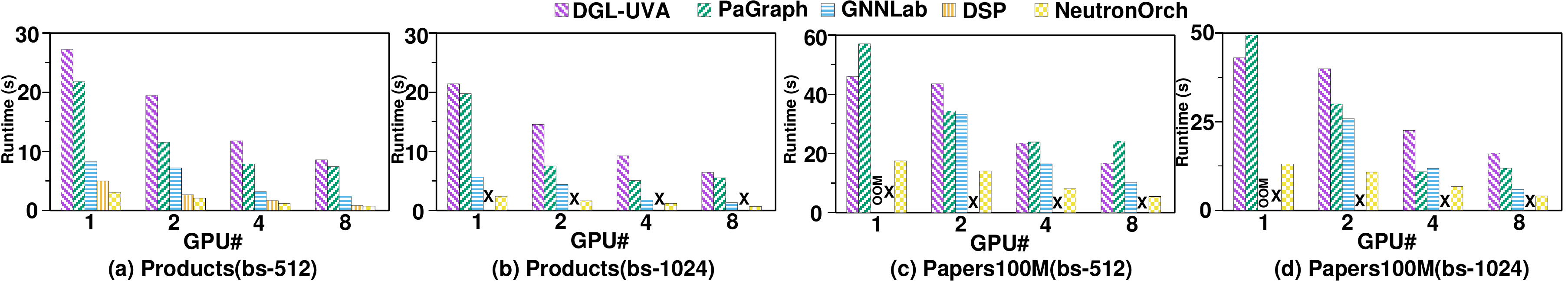}
     \vspace{-0.4cm}
    \caption{Per-epoch runtime of NeutronOrch and other systems under multi-GPU environment. (bs denotes batch size, ``x'' denotes illegal memory access, and ``OOM'' denotes out of memory).}
   \vspace{-0.2cm}
  \label{fig:mulit-GPU}
\end{figure*}

(4) Comparison with DGL-UVA \cite{DGL_CORR_2019}: \system achieves a speed-up ranging from 1.81$\times$ (for GCN on Orkut) to 9.18$\times$ (for GCN on PR) compared to DGL-UVA. DGL-UVA supports accessing graph topology and features via the zero-copy transfer engine \cite{PYTDIRECT_VLDB_2021}. We categorize DGL-UVA's gather step on the CPU since it keeps all the features in the CPU memory without feature caching. 
% The zero-copy transfer engine helps the GPU accelerate the sampling step without saving the graph topology in GPU memory and eliminates the feature collection processing in the gathering step. 
Compared to saving the graph topology in the GPU for sampling, DGL-UVA saves GPU memory while introducing access latency between CPU and GPU. On the other hand, DGL-UVA has a larger communication compared to \system because it transfers all features needed for training in every iteration. 
% as discussed in Section \ref{sec3}, performing entire sample and train steps on the GPU can result in resource contention and prolonged execution time. 

(5) Comparison with GAS \cite{GNNautoscale_icml_2021}: \system achieves a speed-up ranging from 7.08$\times$ (for GAT on Reddit) to 11.05$\times$ (for GAT on PR) compared to GAS. This is because GAS computes historical embeddings for all vertices and transfers them back to the CPU memory. Although GPU memory is saved, additional overhead is incurred due to frequent CPU-GPU communication. GAS also faces CPU memory limitations on graphs with many vertices, as it needs to store embeddings for all vertices across every layer.
In contrast, NeutronOrch selectively computes historical embeddings for frequently accessed vertices on the CPU, achieving efficient historical embedding reuse while reducing memory overhead. 
%This allows for a higher reuse probability with minimal CPU load.
Furthermore, NeutronOrch achieves higher accuracy with strictly bounded staleness than GAS, which arbitrarily reuses historical embeddings within an epoch (see more details in Section \ref{train_convergence}).

% In general, \system achieves its performance over competitors through two primary aspects. First, by employing the hotness-aware layer-based task orchestrating method and a super-batch pipelined training, \system achieves balanced utilization of heterogeneous resources and higher CPU and GPU utilization compared to step-based task orchestrating methods. Second, by utilizing hotness-aware embeddings reusing with bounded staleness, \system effectively reduces the overall computation and transfer data volume while ensuring model accuracy.
% \wqg{ DO WE HAVE SUCH A METHOD? I dont think you clearly explain why DGL-UVA is worse than us, the reason seems quite general and not convenience\# and lacks a GPU-based caching method. }

\vspace{-0.2cm}
\subsection{Multi-GPU Performance}
% \Paragraph{Overall comparison.}
We conduct a comparative analysis of NeutronOrch, PaGraph \cite{PAGRAPH_SOCC_2020}, DGL-UVA \cite{PYTDIRECT_VLDB_2021}, GNNLab \cite{GNNlab_EUROSYS_2022}, and DSP \cite{DSP_PPOPP_2023} to evaluate the scalability by varying the number of GPUs used in training. Figure \ref{fig:mulit-GPU} shows the results of training GraphSAGE against two real-world datasets with different numbers of GPUs. 
\system consistently demonstrates superior performance over the baselines with different batch sizes and different numbers of GPUs.
Compared to DGL-UVA \cite{PYTDIRECT_VLDB_2021} and PaGraph \cite{PAGRAPH_SOCC_2020}, NeutronOrch achieves on average 6.33$\times$ and 5.20$\times$ speedups. The performance of DGL-UVA and PaGraph is limited by extensive CPU-GPU communication and inefficient CPU sampling. In contrast, NeutronOrch minimizes CPU-GPU communication by reusing the historical embeddings and adaptively adjusting the workload between CPU and GPUs.
Compared to GNNLab \cite{GNNlab_EUROSYS_2022} and DSP \cite{DSP_PPOPP_2023}, NeutronOrch achieves on average 2.28$\times$ and 1.36$\times$ speedups. 
GNNLab and DSP deploy all steps (sample-gather-train) on the GPU and leave the CPU idle. When the number of GPUs decreases, GPU memory contention makes the benefit of their caching method decrease. 
% On Products with 4 and 8 GPUs, DSP and GNNLab show similar performance because their cache design effectively reduces CPU-GPU communication. 
In addition, when handling large-scale graphs (Papers100M), both DSP and GNNLab report memory errors due to memory exhaustion. %Notably, DSP even fails to operate on small graphs when using a batch size of 1024. 
NeutronOrch effectively trains large-scale GNNs by offloading computations to the CPU, reducing both CPU-GPU communication and GPU memory overhead.

% \Paragraph{Scaling Performance.}
% The execution time for all systems decreases as more GPUs are added. Specifically, when the number of GPUs increases from 1 to 8, the runtimes of \system, PaGraph, DGL-UVA, GNNLab, and DSP decrease by 3.54$\times$, 3.01$\times$, 2.97$\times$, 4.34$\times$, and 6.07$\times$, respectively. 
% % The scaling performance of DGL-UVA is limited by extensive CPU-GPU communication. In our GPU server, the 8 GPUs are connected to the CPU through 4 PCIe switches, with each two GPUs sharing the same PCIe link to fetch data from the CPU, causing bandwidth contention during training.
% % The scaling performance of PaGraph is limited by inefficient CPU sampling. 
% NeutronOrch, PaGraph, and DGL-UVA fail to achieve a speedup greater than 4$\times$ due to their reliance on CPU-GPU communication for data transfer. In our GPU server, the 8 GPUs are connected to the CPU through 4 PCIe switches, with each two GPUs sharing the same PCIe link to fetch data from the CPU, causing bandwidth contention during training.
% Among these three systems, NeutronOrch achieves better scaling performance by reducing CPU-GPU communication through historical embedding reuse and feature caching. 
% GNNLab and DSP achieve better speedup in successful runs. Nevertheless, their design, which involves caching vertex features and the entire graph topology, limits their ability to handle large-scale graphs. NeutronOrch consistently exhibits high efficiency in all scenarios thanks to the memory-efficient task orchestrating method.

\begin{figure}
  \centering
   % \vspace{-0.1in}
  \includegraphics[width=8cm,page={1}]{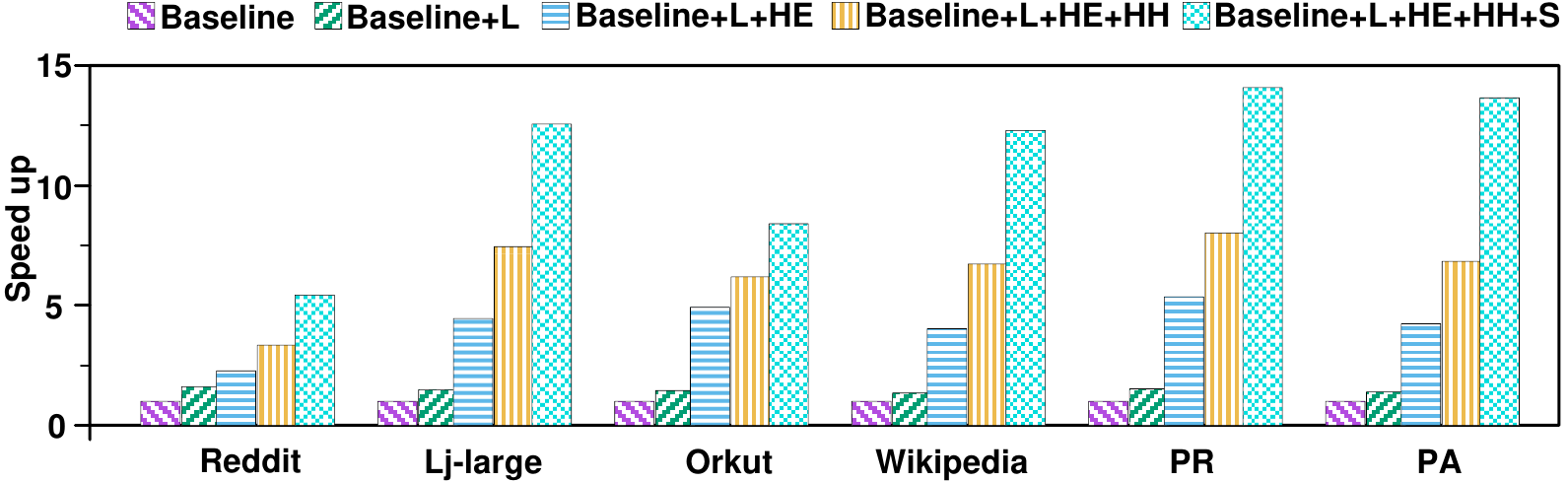}
     \vspace{-0.2cm}
  \caption{Performance gain analysis. L for the layer-based task orchestrating, HE for the hotness-aware embedding reusing, HH for the hybrid hot vertices processing, and S for the super-batch pipelined training.}
   \vspace{-0.2cm}
  \label{fig:breakdown}
\end{figure}

%In the baseline+L configuration, the entire bottom layer is executed in the CPU. 
%1.477942079	2.844361375	 1.642410566
% \wqg{I don't know what is the basic optimization, also, I dont know what is the baseline}
% \wqg{In NeutronStar and HytGraph, we have mature words to introduce the perf. gain analysis, but you still invente a set of  new scentences}
% \vspace{-0.2cm}
\subsection{Performance Gain Analysis}
In this section, we analyze the performance gain of layer-based task orchestrating (L), hotness-aware embedding reusing (HE), hybrid hot vertices processing (HH), and super-batch pipelined training (S) on the GCN model with six datasets. 
We start from a baseline with \system’s codebase for a fair comparison and gradually integrate four optimizations.
The baseline is a step-based task orchestrating method that employs GPU-based graph sampling, GPU-based training, and CPU-based gather step.
Figure \ref{fig:breakdown} shows the normalized speedups. 
%\sanzo{replace with: layer-based task orchestrating}
The layer-based task orchestrating aggregates and updates all vertex features on the CPU and transfers the vertex embedding to the GPU for computation, significantly reducing CPU-GPU communication. 
It can achieve an average speedup of 1.52$\times$ compared to the baseline. 
The hotness-aware embedding reusing optimizes the layer-based task orchestrating of \system, significantly reducing CPU computation. It provides an additional average speedup of 1.96$\times$ over the baseline+L. 
The hybrid hot vertices processing further reduces CPU-GPU communication by effectively balancing CPU and GPU resource utilization. 
It provides an additional average speedup of 1.53$\times$ over the baseline+L+HE. 
Finally, the super-batch pipeline design provides an additional average speedup of 1.61$\times$ over the baseline+L+HE+HH. This design enables the overlap of computations between the CPU and GPU, reducing overall execution time substantially.

% \wqg{Why place the following paragraph here? ? SEE COMMENTS}
%1这个东西看起来应该出现在5.2 而不是这里。
%2）这个东西全是废话，没看出来要说明什么。
%3这种毫无营养的废话会消磨耐心

%三个算法 GCN GraphSage，GAT（pagraph gnnlab不支持GAT） 六个数据集（待定，可以减为4-5个，考虑性能挑选），对比四个计算模式的代表系统

% 固定T在GPU，S D在CPU或GPU四种情况 4 - 4 
%1.GPU： S  T     CPU: D        GNNlab 关掉cache （GNNlab使用显式拷贝，关闭cache即可）
%2.GPU： D  T     CPU: S       Pagraph GNNlab CPU sampling （基于cache策略的 Dataloading，同时采样在CPU）
%3.GPU： S  D  T  CPU: idle     GNNlab 单GPU （性能最佳，打败这个即可） 
%4.GPU： T        CPU: S  D     DGL 0.6版本 (比较朴素的实现，最垃圾)
%5.our system
%整体性能对比，100个eopch，平均一个epoch的时间. 三个算法
% \zyf{don't understand what degree, presample, HER are.}

\begin{figure}[t]
  \centering
  \includegraphics[width=8cm,page={1}]{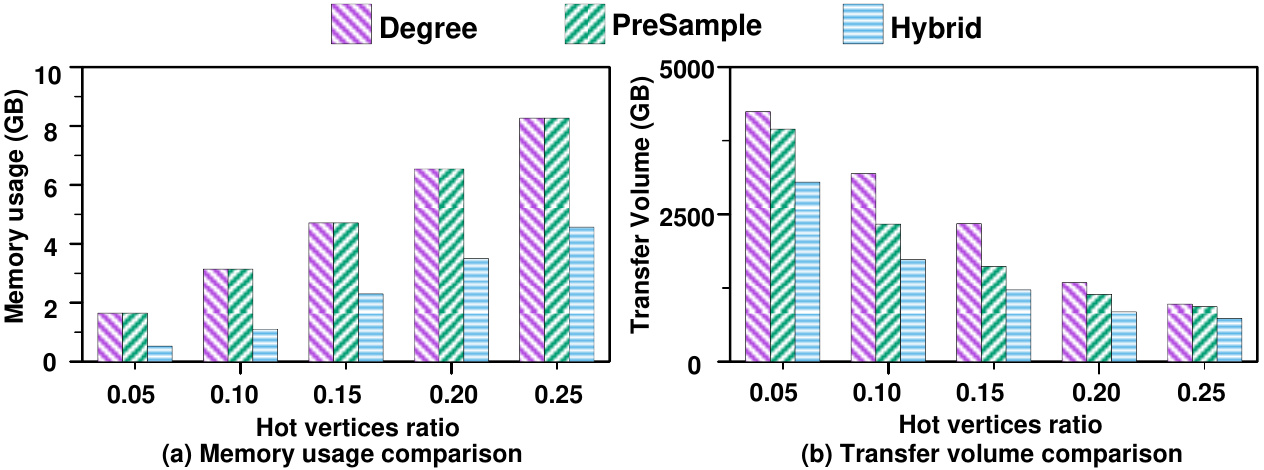}
  \vspace{-0.2cm} %调整图片与上文的垂直距离
  \caption{The memory consumption and transfer volume comparison among different caching policies.}
  \label{fig:hot vertices}
   \vspace{-0.2cm}
\end{figure}

% \wqg{Do we have a outline? I dont see any intuition why evaluating the mentioned things needs implementing caching.}
% \wqg{You should: First, we.... ; Second We .... On the otherhand, the following discussion has very very very very LOW information entropy. Simply repeat our design, nothing interesting, nothing useful, And seems have no "analysis"; See COMMENTS}
% %能不能把最基本的图的顺序和文字顺序做到一致？在caption里面把信息都补足

%节省GPU内存，GPU内存保存两个superbatch内版本的embedding，当前版本和上一版本。
\vspace{-0.1cm}
\label{sec:hot-layer}
\subsection{Benefit of Hotness-Aware Layer-Based Task Orchestrating}

We next explore how the hotness-aware layer-based task orchestrating method impacts the CPU-GPU communication, GPU memory consumption, and GPU training time.

We implement the degree-based cache policy (Degree) \cite{PAGRAPH_SOCC_2020} and the pre-sample-based cache policy (PreSample) \cite{GNNlab_EUROSYS_2022} in \system and compare them with hybrid hot vertices processing (Hybrid). We conduct experiments using the GCN model on the Wikipedia dataset. Figure \ref{fig:hot vertices} (a) illustrates that the Hybrid in \system leads to an average reduction of 55.1\% in GPU memory consumption compared to static cache policies, as embeddings generally have smaller dimensions than features.
On the other hand, although the embedding cache needs to be updated dynamically, each cache hit saves more raw feature transfers.
Under different hot vertices ratios, the average transfer volume using the Hybrid method is 63.2\% and 75.8\% of the static caching strategies Degree and PreSample, respectively.

\begin{figure}[!t]
  \centering
  \vspace{-0.2cm}
  \includegraphics[width=8cm,page={1}]{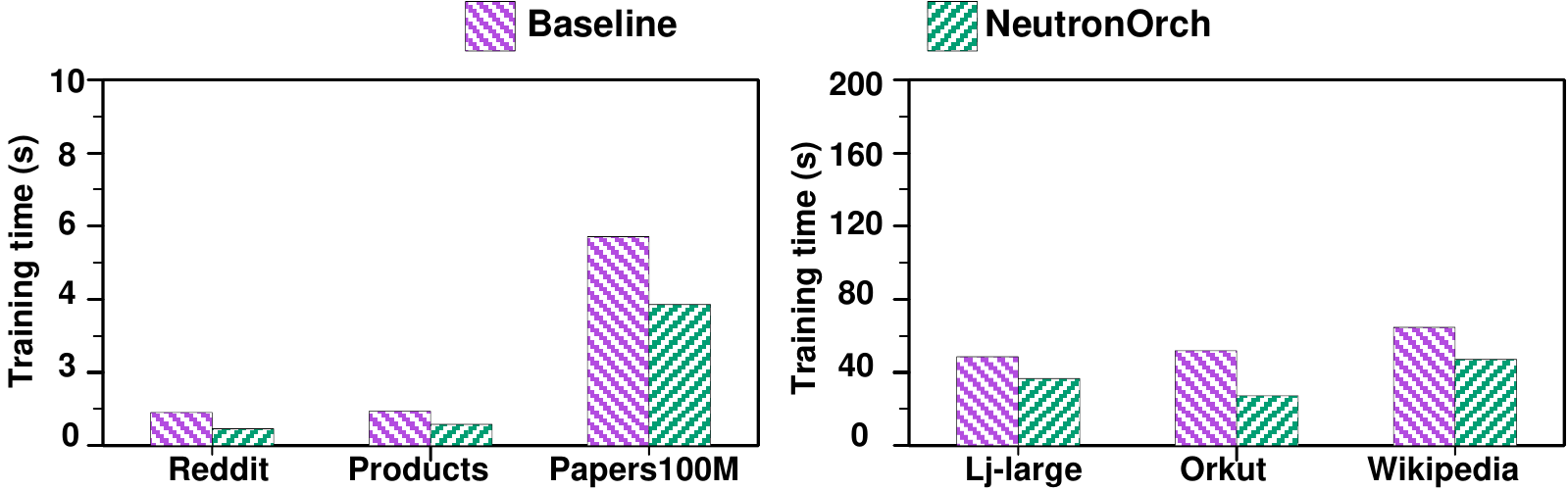}
  \vspace{-0.4cm} %调整图片与上文的垂直距离
  \caption{Average GPU training time in one epoch of Baseline and \system.}
  \label{fig:computation volume}
   \vspace{-0.2cm}
\end{figure}

\begin{table}[!t]
% \vspace{-0.1in}
\caption{Per-epoch runtime of different systems with different model depths (GCN).}
\scriptsize
\vspace{-0.1in}
\label{table:vary_layer}
\centering
\begin{tabular}{c|c|c|c|c|c|c} 
    % \toprule
    \hline
    % \cline{4}
    \multirow{2}{*}{\textbf{Systems}} & \multicolumn{3}{c|}{\textbf{Products}} & \multicolumn{3}{c}{\textbf{Wiki}} \\ 
    \cline{2-7}
    & 3-layer & 4-layer & 5-layer & 3-layer & 4-layer & 5-layer\\ 
    \hline
    DGL & 28.1 & 55.4 &  114.8& 1669.1 & OOM  & OOM \\ 
    \hline 
    PaGraph & 20.1 & 45.0 & 78.0  & 693.2 & OOM  & OOM \\ 
    \hline 
    DGL-UVA & 21.8 & 41.6 & 71.3& 911.7 & 1782.3 & OOM \\ 
    \hline 
    GNNLab & 5.66 & 10.7 & 22.9 & 384.2 & OOM  & OOM \\ 
    \hline 
    GAS & 26.2 & 30.9 & 35.2 & OOM  & OOM  & OOM  \\ 
    \hline 
    % DSP & OOM & OOM & OOM & OOM  & OOM  &OOM \\ 
    % \hline 
    NeutronOrch & \textbf{2.33} & \textbf{4.58} &\textbf{10.2} & \textbf{221.1} & \textbf{483.4} & \textbf{1852.3}\\ 
    \hline
\end{tabular}
\vspace{-0.1in}
\end{table}

\begin{table}[!t]
\vspace{-0.1in}
\caption{Per-epoch runtime of different systems with different batch sizes. (3-layer GCN)}
\scriptsize
\vspace{-0.1in}
\label{table:vary-batchsize}
\centering
\begin{tabular}{c|c|c|c|c|c|c|c|c} 
    % \toprule
    \hline
    % \cline{4}
    \multirow{3}{*}{\textbf{Systems}} & \multicolumn{4}{c|}{\textbf{Products}} & \multicolumn{4}{c}{\textbf{Wiki}} \\ 
    \cline{2-9}
    &\multicolumn{4}{c|}{Batch size} & \multicolumn{4}{c}{Batch size} \\ 
    \cline{2-9}
    & 256 & 1024 & 4096 & 10000 & 256 & 1024 & 4096 & 10000\\ 
    \hline
    DGL & 35.1 & 28.1 & 11.9& 5.24 & 2104.1 & 1669.1 & 861.9  & OOM \\ 
    \hline 
    PaGraph & 31.6 & 20.1 & 15.5  & 11.9 & 1054.6  & 693.2 & OOM  & OOM\\ 
    \hline 
    DGL-UVA & 30.1 & 21.8 & 8.87 & 4.99 & 1624.5  & 911.7 & 592.4 & 301.6\\ 
    \hline 
    GNNLab & 12.3 & 5.65 &  2.78 & 1.65 & 774.8  & 384.2 & OOM  & OOM \\ 
    \hline 
    GAS & 56.4 & 26.2 & 18.1 & 15.1  & OOM  & OOM  & OOM  & OOM\\ 
    \hline 
    % DSP & OOM & OOM & OOM & OOM  & OOM  &OOM & OOM  & OOM\\ 
    % \hline 
    NeutronOrch & \textbf{4.19}  & \textbf{2.33} & \textbf{1.35} & \textbf{0.71} & \textbf{409.2} & \textbf{221.1} &\textbf{129.9} & \textbf{84.6}\\ 
    \hline
\end{tabular}
\vspace{-0.12in}
\end{table}

On the other hand, the GPU training time is significantly reduced since the CPU-provided hot vertex embeddings can be reused multiple times within the super-batch.
A larger hot vertices ratio facilitates more GPU computation savings. The proportion of hot vertices is proportional to the CPU's computing power. In order to guarantee the bound, the CPU must complete the hot vertices embeddings update of the next super-batch within one super-batch. The proportion of hot vertices that can be supported by different datasets is usually 10\%-30\%. We use the \system with a hot vertices ratio of 0 as a baseline and compare the average GPU training time of one epoch with \system.
Figure \ref{fig:computation volume} presents the observation of GPU training time reduction for GCN algorithms with six datasets. \system reduced GPU training time by an average of 36.5\% on six datasets. 
In addition, the reduction of GPU training time is most significant on Reddit and Orkut, with an average of 48.3\%. This is because they have a larger average degree, which is great for improving hots vertices hits. 

\begin{figure}
  \centering
  \includegraphics[width=8cm,page={1}]{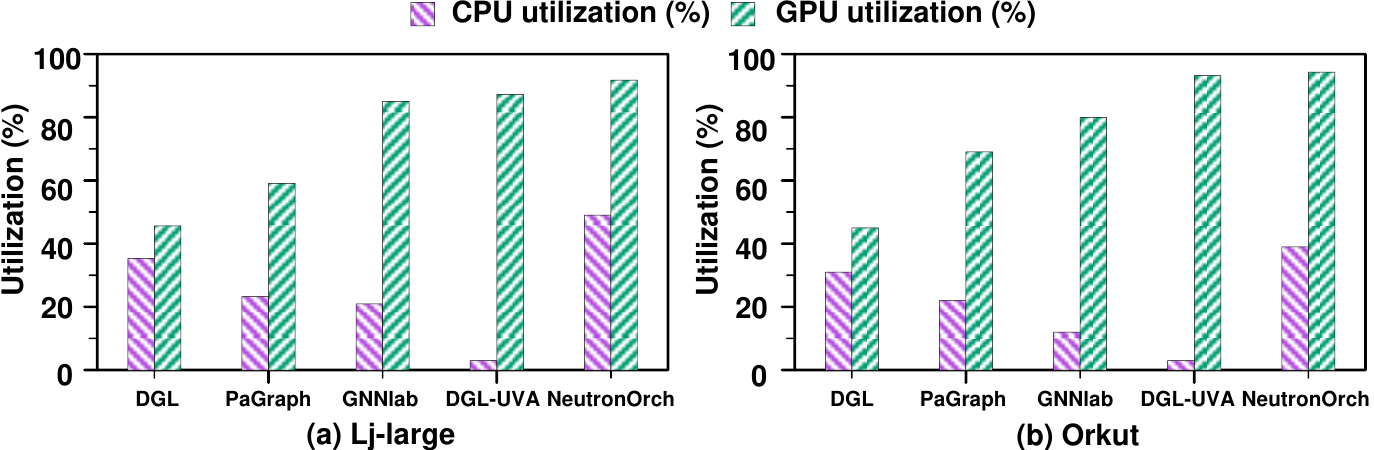}
   \vspace{-0.13in}
  \caption{GPU utilization and CPU utilization comparison.}
   \vspace{-0.4cm}
  \label{fig:uti-exp}
\end{figure}

\vspace{-0.1in}
\subsection{Analysis of Resource Utilization}
% 对比对象 ：1.在本系统中配置四种计算模式 2.our system
%2.统计四种模式和本系统的CPU与GPU计算资源利用率，证明在异构环境下的新任务分配对资源的充分利用
%3.存储资源利用率怎么对比不知道。
%六个数据集，GCN算法，Graphsage算法
We evaluate the utilization of GPU and CPU resources during the training of GCN on Lj-large and Orkut using five systems: DGL, PaGraph, GNNLab, DGL-UVA, and \system. Figure \ref{fig:uti-exp} presents a comparison of GPU and CPU utilization. GPU and CPU utilization were recorded every 100 milliseconds and averaged over ten epochs. \system exhibits superior utilization of both CPU and GPU resources by scheduling training tasks across CPUs and GPUs. DGL and PaGraph have good CPU utilization and poor GPU utilization. This is because performing a complete sampling or gathering step on the CPU improves CPU utilization but causes the GPU to wait. DGL-UVA and GNNLab have poor CPU utilization and good GPU utilization. They have good performance with sufficient GPU resources, but exploiting CPU resources can further improve performance. 
\system achieves both good CPU and GPU utilization, averaging 44.5\% and 92.9\%, respectively.

\subsection{Varying Model Configurations}

\Paragraph{Performance with varying model depths.}
As the model depth increases, the effectiveness of NeutronOrch will not decrease, although NeutronOrch only offloads the lowest-level calculations to the CPU. In sample-based GNN training, the number of vertices exhibits exponential growth across layers \cite{Graphsage_2017,FASTGcn_iclr_2018}. As a result, more than half of the entire training workload comes from the bottom layer. For example, on Wiki dataset with a 3-layer model, the bottom layer has 1.94M vertices, occupying 65\% of the total workload, while the other layers only have 1.04M vertices in total. For the 4-layer model and 5-layer model, the bottom layer occupies 61\% and 59\% of the total workload, respectively. 
We run a GCN model on all systems and report the per-epoch runtime in Table \ref{table:vary_layer}, with the model depth ranging from 3 to 5. For the 3-layer model, NeutronOrch achieves on average 6.43$\times$ speedup over the different systems. For the 4-layer model and 5-layer model, the speedups are 5.84$\times$ and 6.31$\times$, respectively.

\Paragraph{Performance with varying batch sizes.}
To study the effectiveness of NeutronOrch under different batch sizes, we run a 3-layer GCN model on all systems and report the per-epoch runtime in Table \ref{table:vary-batchsize}, with the batch size ranging from 256 to 10000.
For a batch size of 256, NeutronOrch achieves on average 5.64$\times$ speedups over different systems. For batch sizes of 1024, 4096, and 10000, the speedups are 6.43$\times$, 9.85$\times$, and 7.25$\times$, respectively. The setting of batch size will not affect the effectiveness. Furthermore, on large-scale graphs, NeutronOrch's CPU offloading efficiently mitigates the increasing memory overhead when using large batch sizes.

%第一个时间，一整个的bottom layer 的CPU计算 算完给GPU
%第二个时间，只有hot vertice 的CPU计算，GPU按需的拿
%第三个时间，多流的
%依次开启，新的CPU-GPU任务摆放，热点嵌入重用，看性能提升
%六个数据集，GCN算法

%分析性能提升来源
%我们系统的性能优势应该来自于以下三点，重要性按序号排序： 1.更充分的异构资源利用率 2.更少的传输量 3.更小的计算量 

\begin{figure}
  \centering
  \includegraphics[width=8cm,page={1}]{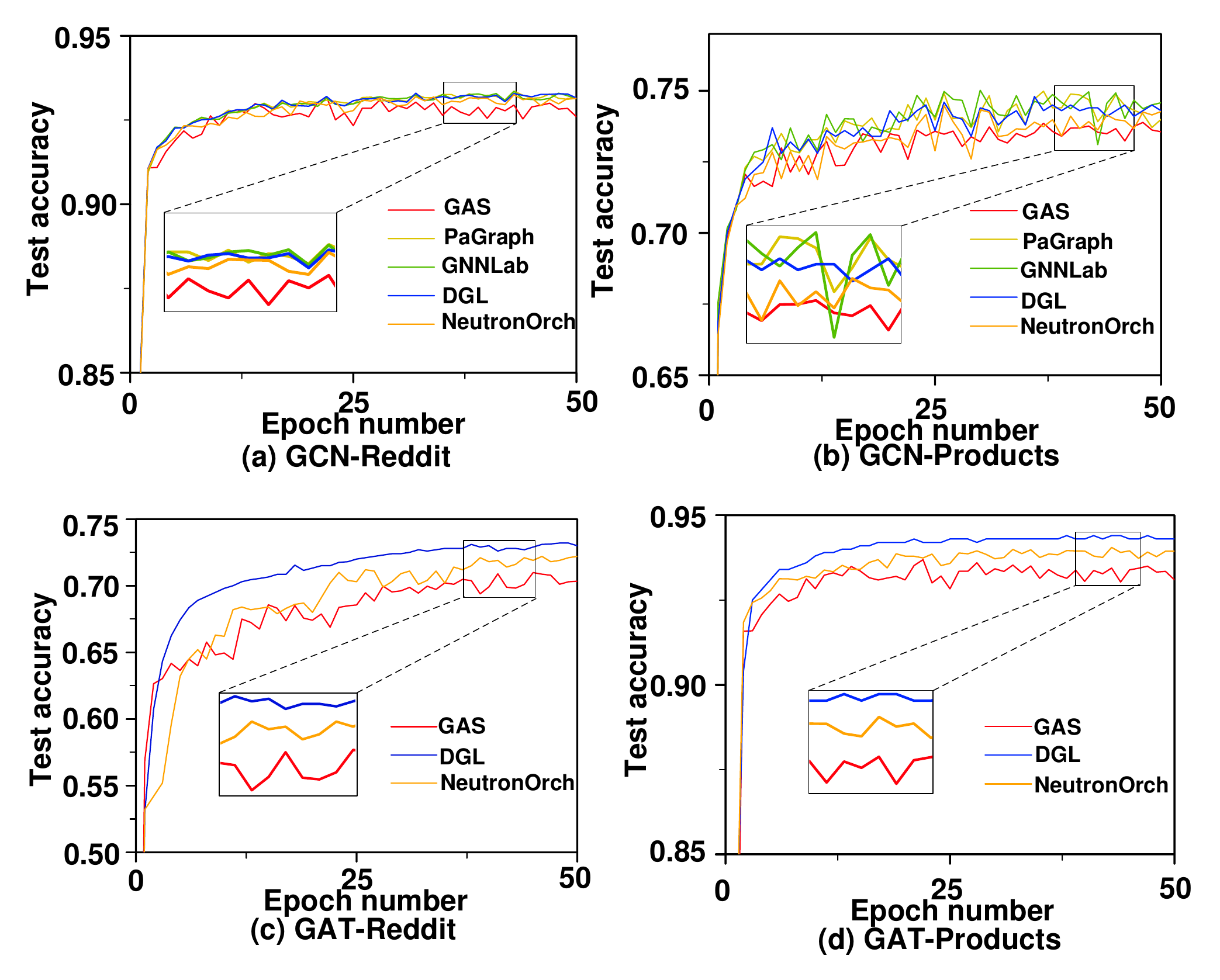}
     \vspace{-0.4cm}
  \caption{Epoch-to-accuracy}
   \vspace{-0.6cm}
  \label{fig:acc}
\end{figure}

\vspace{-0.1cm}
\label{sec:acc}
\subsection{Training Convergence}
\label{train_convergence}
%验证方法的time to accrcuy 
%一个数据集，GCN算法，GAT算法，对比CPU中不同热点个数，对准确率收敛的影响
%证明调节热点个数，可以改善embedding版本陈旧情况
% Benefit from strict control of bounded staleness by super-batch pipelined training, \system optimally utilizes heterogeneous resources without compromising the accuracy of GNNs. Stale bound in the \system depends on the number of batches in the super-batch. In actual training, we find that using a larger bound benefits the overall performance, but the accuracy rise slows down \cite{SANCUS_VLDB_2022,ST_ICML_2018,ssp_2013_nips}. The tradeoff between system performance and model convergence is difficult to choose. Moreover, differences in applications, datasets, implementations, and configurations may affect the practical staleness in the GNN training as well. In our practical tests, we find that setting the number of batches between 4 and 8 provides an acceptable balance between runtime and errors.
We plot the epoch-to-accuracy curve on different systems for both GCN and GAT algorithms.
Benefiting from the strict version control based on super-batch pipelining, \system accelerates GNN training while maintaing high accuracy. As shown in Figure \ref{fig:acc}, compared to other GNN systems that use historical embeddings but do not strictly control versions (e.g., GAS\cite{GNNautoscale_icml_2021}), NeutronOrch achieves higher accuracy due to the ability to achieve smaller cumulative errors across batches. In comparison to systems that do not use historical embeddings\cite{DGL_CORR_2019}, NeutronOrch achieves comparable convergence with an accuracy loss of no more than 1\%.

 \vspace{-0.2cm}
\section{Related Work}
\label{sec6}
\Paragraph{Sample-based mini-batch GNN training.}  GraphSage \cite{Graphsage_2017} first introduces the fixed-number vertex sampling and proposes the general message aggregation methods. Past works mainly focused on solving the sampling and gathering bottleneck in sample-based GNN training. For the sampling step, LazyGCN \cite{lazygcn_2020_neuips} proposes a sample graph reusing method to reduce sampling overhead and maintain accuracy. MariusGNN \cite{Marius++_arxiv_2022} design the DENSE data structure to minimize redundant computation in multi-hop sampling. On the other hand, Nextdoor \cite{Nextdoor_eurosys_2021}, Wholegraph \cite{wholegraph_SC_2022}, and TurboGNN \cite{TurboGNN_TOC_2023} propose the GPU-based sampling method to accelerate the sampling step. For the gathering step, a series of feature and graph caching strategies are proposed. Data tiering \cite{DATATIRING_VLDB_2022}, and Quiver \cite{quiver_2023_arxiv} design various efficient GPU feature cache strategies to minimize CPU-GPU communication. Ducati \cite{ducati_sigmod_2023} further designs the graph topology cache method to fully utilize the GPU memory.
% Pytorch-direct \cite{PYTDIRECT_VLDB_2021} makes GPU threads directly access sparse features in host memory through on-demand zero-copy accesses. 

% \Paragraph{Historical embeddings reusing.}
% Existing works \cite{ST_ICML_2018,SANCUS_VLDB_2022,GNNautoscale_icml_2021,refresh_arxiv_2023} have shown the benefits of historical embedding (HE) reuse for training acceleration without compromising accuracy. VR-GCN \cite{ST_ICML_2018} uses HE in GNN training to reduce the variance and further control the number of nodes to sample. Refresh \cite{refresh_arxiv_2023} uses a combination of gradient-based and staleness criteria to reduce estimation errors. SANCUS \cite{SANCUS_VLDB_2022} reuses HE for boundary vertices to reduce communication in distributed GNN training. These frameworks focus on optimizing scenarios where computation is exclusively deployed on GPUs. In contrast, our work extends HE reuse to the CPU-GPU environment, achieving both high performance and scalability.

% GNNLab \cite{GNNlab_EUROSYS_2022} proposes a pre-sampling-based cache policy to accommodate different graph datasets and sampling algorithms. Data tiering \cite{DATATIRING_VLDB_2022} considers the multi-hop neighbor information of the vertex by the Reverse-PageRank when selecting cache vertices. 

\Paragraph{Task orchestrating in DNN training.} 
Zero-Infinity DeepSpeed (ZID) \cite{ZID_2020} incorporates CPU offload techniques \cite{ZID_1_SC_2020,ZID_2_SC_2021,ZID_ATC_2021} to optimize the DNN training under heterogeneous environments. Although ZID and our design share similarities in scheduling training tasks across CPU and GPUs to improve scalability, their optimization objectives are different.
ZID primarily focuses on reducing the memory consumption of model parameters, reducing memory consumption by offloading model parameters to the CPU. In DNNs, model parameters which consist of dense matrices that can be disjointly partitioned. These slices can be efficiently communicated between the CPU and GPUs because of their dense and regular data access patterns. 
In contrast, NeutronOrch primarily focuses on the efficiency of sample-based GNN training, minimizing communication and computation by offloading historical embedding computation to the CPU. 
Sample-based GNN training involves multiple data preparation stages (sample and gather). It exhibits irregular vertex access patterns due to the inherent complexity of graph structures, which pose new challenges in optimizing the data I/O between CPU and GPUs. ZID didn't address these challenges. However, NeutronOrch effectively resolves these problems by its task orchestrating method, transferring and reusing historical embedding for frequently accessed vertices. 

\vspace{-0.2cm}
\section{Conclusion}
\label{sec7}
We present \system, a scalable and efficient GNN training system that fully utilizes CPU and GPUs. 
\system leverages two key components to achieve its performance, including a hotness-aware layer-based task orchestrating method that combines CPU computation offloading with historical embeddings reuse to optimize computation and communication and a super-batch pipeline training method that utilizes CPU-GPU pipelining to achieve efficient and staleness-bounded version control. 
Our experiments demonstrate that \system efficiently accelerates mini-batch GNN training with an accuracy loss of no more than 1\%.

% for training sample-based mini-batch GNNs under CPU-GPU heterogeneous environments
% The effectiveness of \system and its high performance are contributed by several system components, including a layer-based task orchestrating method, hotness-aware embedding reusing, and super-batch pipeline scheduling.
% on billion-scale graphs using just 4 GPUs by fully utilizing CPU, GPU, and interconnects.

% Compared with existing mini-batch GNN systems DGL \cite{DGL_CORR_2019}, PaGraph \cite{PAGRAPH_SOCC_2020}, GNNAutoScale \cite{GNNautoscale_icml_2021}, GNNLab \cite{GNNlab_EUROSYS_2022}, and DSP \cite{DSP_PPOPP_2023}, the speedups of \system range from 1.20 $\times$ to 4.91$\times$ by achieving a balanced computation resource utilization.

% \begin{acks}
%  This work was supported by the [...] Research Fund of [...] (Number [...]). Additional funding was provided by [...] and [...]. We also thank [...] for contributing [...].
% \end{acks}

%\clearpage
\balance
\bibliographystyle{ACM-Reference-Format}
\bibliography{sample}

\end{document}